\documentclass[10pt,aps,prb,twocolumn,english,superscriptaddress,
               floatfix,longbibliography]{revtex4-2}

\usepackage{graphicx}       
\usepackage{dcolumn}        
\usepackage{bm}             
\usepackage{lmodern}
\usepackage[utf8]{inputenc}
\usepackage{color}
\usepackage{varioref}
\usepackage{float}
\usepackage{units}
\usepackage{mathtools}
\usepackage{enumitem}
\usepackage{physics}
\usepackage{amsmath}
\usepackage{amsthm}
\usepackage{amssymb}
\usepackage{braket}
\usepackage{array}
\usepackage{calc}
\usepackage[table]{xcolor}
\usepackage{multirow}
\usepackage[unicode=true,pdfusetitle,
            bookmarks=true,bookmarksnumbered=false,bookmarksopen=false,
            breaklinks=true,pdfborder={0 0 0},pdfborderstyle={},
            backref=false,colorlinks=true]{hyperref}

\begin{document}

\preprint{APS/123-QED}

\title{Local Defects and the Topology of the Haldane Model}

\author{Vishesh Makwana}
\affiliation{%
  Department of Physics, Technion -- Israel Institute of Technology,
  Haifa 3200003, Israel}

\author{Eric Akkermans}
\affiliation{%
  Department of Physics, Technion -- Israel Institute of Technology,
  Haifa 3200003, Israel}

\begin{abstract}
We investigate the interplay between local defects and topology in the Haldane model within the framework of the tenfold classification. The Haldane model realizes a Chern-insulating phase characterized by an integer topological invariant ($C=\pm 1$) and supports chiral edge states.  Introducing vacancies gives rise to localized states at the defect sites, classified by a $\mathbb{Z}_2$ invariant $\nu = C\cdot m\,\mathrm{mod}\,2$, where $m=N_A - N_B$ is the net sublattice imbalance of the vacancy configuration: an odd imbalance hosts a protected zero-energy mode, whereas an even imbalance does not.  We identify three independent experimental signatures that distinguish these topological defect states from trivial (adatom) defects. First, vacancy-induced states exhibit characteristic dislocations in their wavefunction profiles that track the phase winding associated with the defect. Second, a
fractional charge of $e/2$ accumulates at vacancy sites, while no such charge appears at adatoms. Third, the probability current circulating around a vacancy-induced state flows in the \emph{opposite} direction to that of chiral edge states, in
direct analogy with the current reversal produced by a vortex in a $p$-wave superconductor. All three signatures are in quantitative agreement with the $\mathbb{Z}_2$ prediction.
\end{abstract}

\maketitle

\section{Introduction}
\label{sec:intro}

In recent decades, topology has emerged as a powerful organizing principle in
condensed matter physics.  Unlike conventional phases of matter, which are
classified by local order parameters and symmetry breaking, topological phases
are characterized by global properties of the Hamiltonian that remain invariant
under smooth deformations.  These properties give rise to robust physical
phenomena that are insensitive to local perturbations, provided that certain
symmetries are preserved and the bulk energy gap remains open.  A striking
example is the integer quantum Hall effect~\cite{Klitzing1980}, where the Hall
conductance is quantized and determined by a topological invariant known as the
Chern number~\cite{Thouless1982,KOHMOTO1985,Hatsugai1993}.

A systematic framework for classifying topological phases of non-interacting
fermionic systems is provided by the tenfold
classification~\cite{Altland1997,Kitaev2009,Schnyder2008}. This scheme organizes Hamiltonians according to the presence or absence of three fundamental symmetries: the anti-unitary time-reversal and particle-hole symmetries, and the unitary chiral symmetry.
For each symmetry class and spatial dimension, the classification predicts
whether nontrivial topological phases can exist and identifies the associated
invariant ($\mathbb{Z}$ or $\mathbb{Z}_2$).

Beyond bulk phases, topology can also manifest in the presence of
defects~\cite{Teo2010}.  Topological defects host localized bound states
whose existence is determined by the symmetry class and codimension of the
defect.  These states provide experimentally accessible signatures of topology,
often appearing as zero-energy modes bound to the defect
core~\cite{Ugeda2010}.  Vacancies in graphene are a paradigmatic
example~\cite{Kelly1998,Ugeda2010,Ovdat2017,Pereira2006,Pereira2008,
Amara2007,Palacios2008,Dutreix2013,Goft2023,Godsil2001,Faccio2010,
Yang2018,abulafia_wavefronts_2023}, providing a natural platform to study
defect-induced topological phenomena.

In this paper, we study a system in which \emph{both} bulk and defect topology
appear simultaneously.  The bulk Haldane model possesses a nontrivial
$\mathbb{Z}$ invariant $C=\pm 1$, while point vacancies within the same
system exhibit a $\mathbb{Z}_2$ classification. The central result is that
these two levels of topology are not independent: they are coupled by the
relation
\begin{equation}
  \nu = C\cdot m\;\mathrm{mod}\;2,
  \label{eq:central}
\end{equation}
where $m=N_A - N_B$ is the net sublattice imbalance of the vacancy
configuration.  This bulk-defect coupling formula is the organizing principle
of our paper.  We derive it analytically via the Weyl symbol of the Hamiltonian
(Appendix~\ref{app:symbol}) and the Chern-Simons invariant
(Appendix~\ref{app:z2}), and we verify it through three independent numerical
diagnostics: wavefunction dislocations, fractional charge, and current
reversal.

The remainder of the paper is organized as follows.  Section~\ref{sec:bulk}
introduces the Haldane model and its bulk topology.
Section~\ref{sec:defects} defines the vacancy and adatom defects and derives
Eq.~\eqref{eq:central}.  Sections~\ref{sec:modes}--\ref{sec:currents}
present the three numerical signatures.
Section~\ref{sec:discussion} discusses the analogy with $p$-wave
superconductors and experimental implications.
Section~\ref{sec:conclusion} concludes.

\section{The Haldane Model}
\label{sec:bulk}

We start with the tight-binding Hamiltonian of the Haldane
model~\cite{haldane1988},
\begin{equation}
\begin{aligned}
  \mathcal{H}_0 =
    -t_1 \sum_{i,\bm{\delta}} a^\dagger_{\bm{R}_i} b_{\bm{R}_i+\bm{\delta}}
    &- t_2 \sum_{i,\bm{\beta}} e^{i\phi}\,
       a^\dagger_{\bm{R}_i} a_{{\bm{R}_i}+\bm{\beta}} \\
    &- t_2 \sum_{i,\bm{\beta}} e^{-i\phi}\,
       b^\dagger_{\bm{R}_i} b_{{\bm{R}_i}+\bm{\beta}}
    + \mathrm{h.c.}
    \label{eq:haldane}
\end{aligned}
\end{equation}
where $t_1$ and $t_2$ denote the nearest-neighbor and next-nearest-neighbor
hopping amplitudes, respectively, and $\phi$ is the Peierls phase associated
with the next-nearest-neighbor hopping, chosen such that the total flux through
each unit cell vanishes.  The operators $a_i^\dagger$ ($a_i$) and
$b_i^\dagger$ ($b_i$) create (annihilate) fermions on the $A$ and $B$
sublattices at site $\boldsymbol{R}_i$, respectively.

\begin{figure}[b]
\centering
\includegraphics[width=0.28\linewidth]{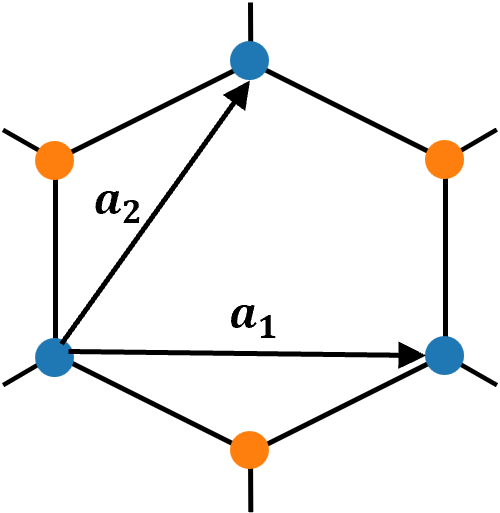}
\hfill
\includegraphics[width=0.30\linewidth]{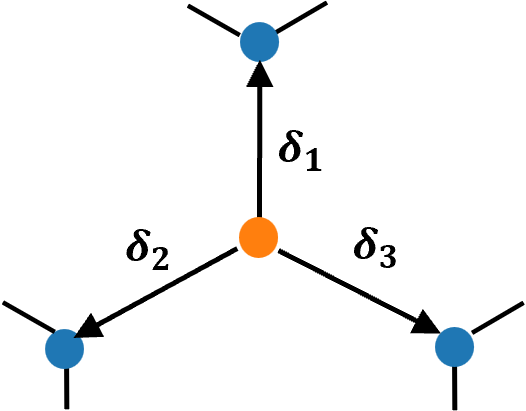}
\hfill
\includegraphics[width=0.28\linewidth]{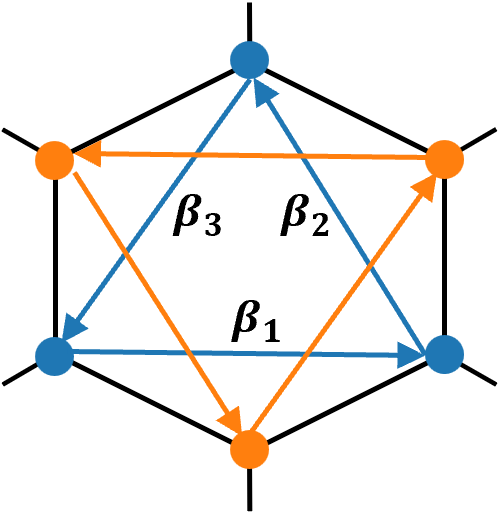}
\caption{Lattice vectors of the honeycomb lattice.  Left: primitive lattice
  vectors $\boldsymbol{a}_{1,2}$.  Center: nearest-neighbor vectors
  $\boldsymbol{\delta}_{1,2,3}$.  Right: next-nearest-neighbor vectors
  $\boldsymbol{\beta}_{1,2,3}$.}
\label{fig:halvec}
\end{figure}

The geometry of the honeycomb lattice is specified by the primitive lattice vectors
\begin{equation}
    \boldsymbol{a}_1 = \sqrt{3}a\,\hat{x},
    \qquad
    \boldsymbol{a}_2 = \frac{\sqrt{3}a}{2}\,\hat{x}
    + \frac{3a}{2}\,\hat{y}.
\end{equation}
The three nearest-neighbor vectors connecting sites on opposite sublattices are
\begin{equation}
\boldsymbol{\delta}_1 = a\,\hat{y}, \qquad
\boldsymbol{\delta}_{2,3}
=
\mp \frac{\sqrt{3}a}{2}\,\hat{x}
-\frac{a}{2}\,\hat{y}.
\end{equation}
The next-nearest-neighbor vectors, which connect sites on the same sublattice are
\begin{equation}
    \boldsymbol{\beta}_1 = \sqrt{3}a\,\hat{x},
    \qquad
\boldsymbol{\beta}_{2,3}
=
-\frac{\sqrt{3}a}{2}\,\hat{x}
\pm \frac{3a}{2}\,\hat{y}.
\end{equation}
A schematic illustration of these vectors is shown in Fig.~\ref{fig:halvec}.

Performing a Fourier transform, the Hamiltonian becomes
$\mathcal{H}_0 = \int_\mathrm{BZ} d\boldsymbol{k}\;
\psi^\dagger_{\boldsymbol{k}} H_0(\boldsymbol{k}) \psi_{\boldsymbol{k}}$,
with the $2\times 2$ Bloch Hamiltonian
\begin{equation}
  H_0(\boldsymbol{k}) =
    \begin{pmatrix}
      t_2 g(\boldsymbol{k},\phi) & t_1 f(\boldsymbol{k}) \\
      t_1 f^*(\boldsymbol{k})   & t_2 g(\boldsymbol{k},-\phi)
    \end{pmatrix},
\end{equation}
where $\psi_{\boldsymbol{k}} = (\psi_{a,\boldsymbol{k}},\,
\psi_{b,\boldsymbol{k}})^T$, and
\begin{align}
  f(\boldsymbol{k})
  &\equiv
  -\!\left(
    e^{i\boldsymbol{k}\cdot\boldsymbol{\delta}_1}
    + e^{i\boldsymbol{k}\cdot\boldsymbol{\delta}_2}
    + e^{i\boldsymbol{k}\cdot\boldsymbol{\delta}_3}
  \right), \\
  g(\boldsymbol{k},\phi)
  &\equiv
  -2\Bigl(
    \cos(\boldsymbol{k}\cdot\boldsymbol{\beta}_1+\phi)
    + \cos(\boldsymbol{k}\cdot\boldsymbol{\beta}_2+\phi)
    \nonumber\\
  &\qquad\qquad
    + \cos(\boldsymbol{k}\cdot\boldsymbol{\beta}_3+\phi)
  \Bigr).
\end{align}
The corresponding energy spectrum is given by, 
\begin{equation}
    \epsilon(\boldsymbol{k}) = \pm\sqrt{t_1^2|f|^2 +
  \tfrac{t_2^2}{4}(g(\boldsymbol{k},\phi)-g(\boldsymbol{k},-\phi))^2},
\end{equation}
which exhibits a finite band gap for $\phi\neq 0,\pm\pi$.  The low-energy Dirac
points lie at $\boldsymbol{K} = -\boldsymbol{K'} =
(\tfrac{4\pi}{3\sqrt{3}a},0)$.  The Chern number of the lower band is
$C=\mathrm{sign}(\sin\phi)$, so $C=+1$ for $\phi=\pi/2$.

\begin{figure}[t]
\centering
\includegraphics[width=0.53\linewidth]{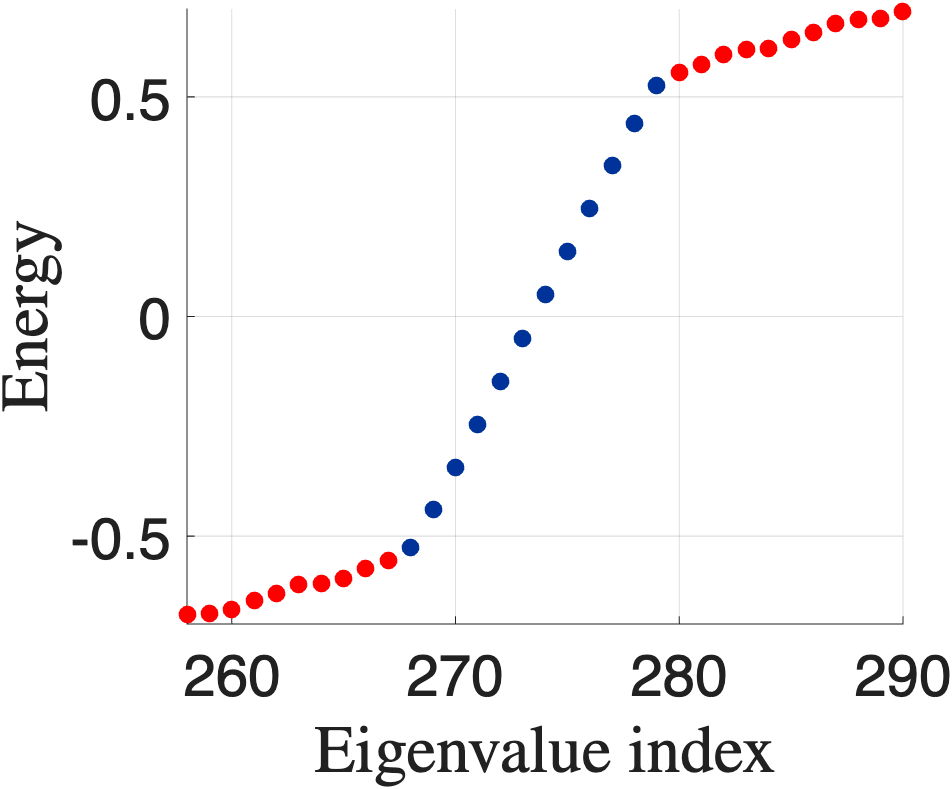}
\hfill
\includegraphics[width=0.42\linewidth]{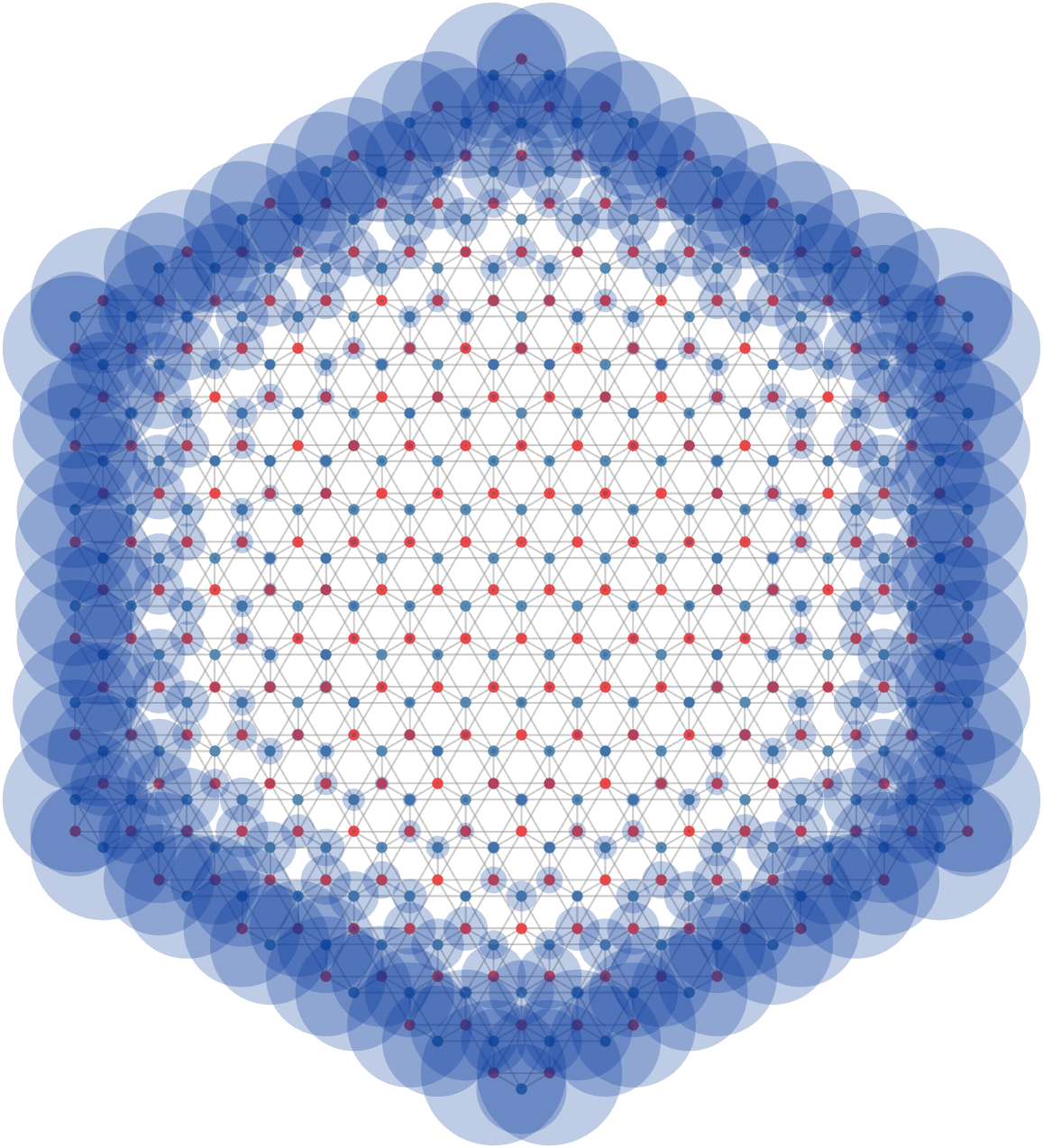}
\caption{Clean Haldane model ($t_1=1$, $t_2=0.1$, $\phi=\pi/2$).  Left:
  energy spectrum showing edge states (blue) inside the bulk gap and bulk modes
  (red) outside.  Right: spatial probability density $|\psi|^2$ of a
  representative edge state (highlighted blue in the spectrum), confirming
  localization along the system boundary.}
\label{fig:haledge}
\end{figure}

By the bulk-boundary correspondence, the nontrivial Chern number implies the
existence of gapless chiral edge states.  This is confirmed numerically in
Fig.~\ref{fig:haledge}: edge states appear inside the bulk gap (blue), while
bulk modes lie outside (red), and the edge-state wavefunction is localized
along the boundary.  All numerical results below use parameters $t_1=1$,
$t_2=0.1$, and $\phi=\pi/2$ unless stated otherwise.

\section{Point Defects and Topological Classification}
\label{sec:defects}

\begin{figure}[b]
\centering
\includegraphics[width=0.49\linewidth]{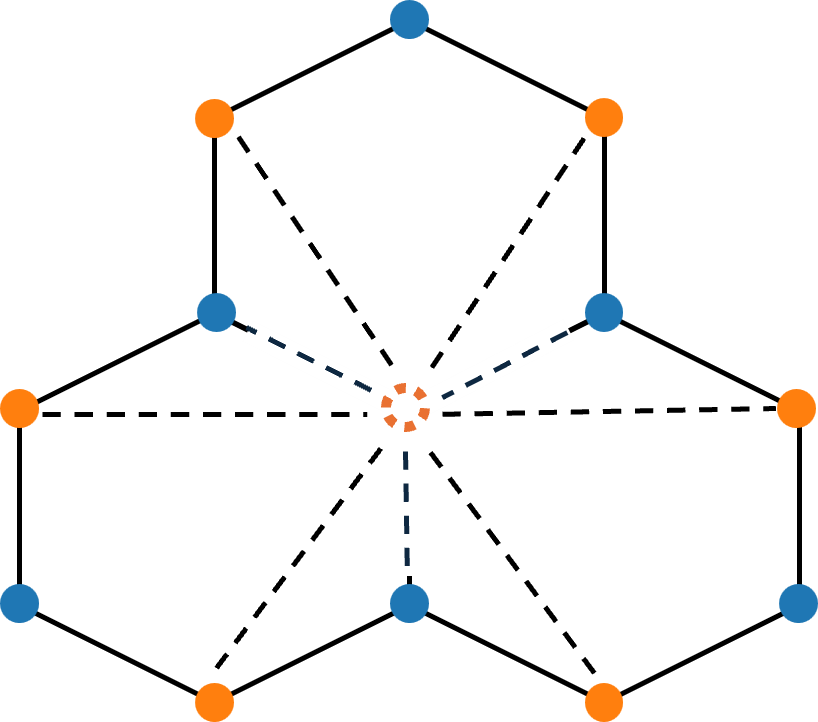}
\hfill
\includegraphics[width=0.49\linewidth]{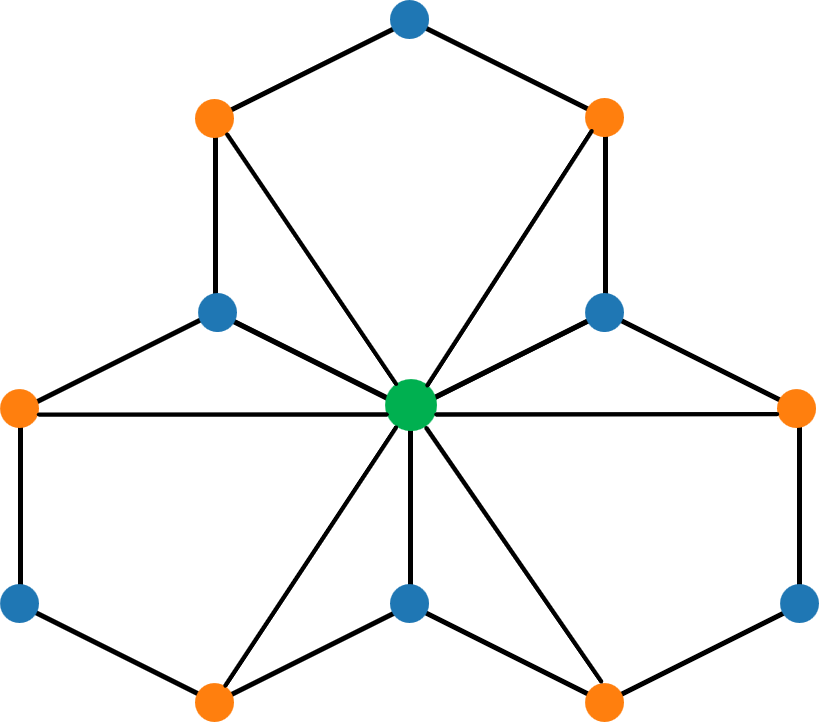}
\caption{Point defects in the Haldane model.  Left: a vacancy is created by
  removing a lattice site along with all its hopping bonds (dotted lines).
  Right: an adatom is introduced as an on-site potential at a lattice site
  (green).}
\label{fig:point_defect}
\end{figure}

We consider two types of point defects.  A \emph{vacancy} is modeled by
removing a single atom together with all hopping bonds connected to it.  An
\emph{adatom} is modeled by adding an on-site energy $\epsilon_0 \to \infty$
at one lattice site, effectively decoupling it without removing hopping
bonds.  Figure~\ref{fig:point_defect} illustrates both cases.

\subsection{Vacancy potential and Weyl symbol}

For a vacancy on the $A$-sublattice at $\boldsymbol{R}_0$, the perturbation
to the Hamiltonian is
\begin{subequations}
\label{eq:vac_real}
\begin{align}
  \mathcal{V}_\mathrm{vac}^{(1)}(\boldsymbol{R}_0)
    &= t_1 \sum_i \Bigl(
         a^\dagger_{\boldsymbol{R}_0}\,
         b_{\boldsymbol{R}_0+\boldsymbol{\delta}_i}
       + \mathrm{h.c.}\Bigr),
  \label{eq:vac_real_nn} \\[4pt]
  \mathcal{V}_\mathrm{vac}^{(2)}(\boldsymbol{R}_0)
    &= t_2 \sum_i
    \begin{aligned}[t]
      \Bigl(
      e^{i\phi}\,
       a^\dagger_{\boldsymbol{R}_0}\,
       &a_{\boldsymbol{R}_0+\boldsymbol{\beta}_i}
       + e^{-i\phi}\,
       a^\dagger_{\boldsymbol{R}_0}\,
       a_{\boldsymbol{R}_0-\boldsymbol{\beta}_i}
       \\
      &+ \mathrm{h.c.}
      \Bigr),
    \end{aligned}
  \label{eq:vac_real_nnn}
\end{align}
\end{subequations}
so that $\mathcal{V}_\mathrm{vac} = \mathcal{V}_\mathrm{vac}^{(1)} +
\mathcal{V}_\mathrm{vac}^{(2)}$, where \eqref{eq:vac_real_nn} collects the
nearest-neighbor bonds removed by the vacancy and \eqref{eq:vac_real_nnn}
collects the next-nearest-neighbor bonds.
In momentum space this becomes
\begin{equation}
\begin{aligned}
  V(\boldsymbol{R}_0)
    = \frac{t_1}{N} \sum_{\boldsymbol{k},\boldsymbol{k'}}
       e^{i(\boldsymbol{k'}-\boldsymbol{k})\cdot\boldsymbol{R}_0}
       f(\boldsymbol{k'})&\, a^\dagger_{\boldsymbol{k}} b_{\boldsymbol{k'}}
  \\
    + \frac{t_2}{N} \sum_{\boldsymbol{k},\boldsymbol{k'}}
       e^{i(\boldsymbol{k'}-\boldsymbol{k})\cdot\boldsymbol{R}_0}
       g(\boldsymbol{k'},\phi)&\, a^\dagger_{\boldsymbol{k}} a_{\boldsymbol{k'}}
    + \mathrm{h.c.},
  \label{eq:vac_pot}
\end{aligned}
\end{equation}
where $N$ is the number of unit cells.  The phase factor
$e^{i(\boldsymbol{k'}-\boldsymbol{k})\cdot\boldsymbol{R}_0}$ mixes different
momenta and, in particular, induces intervalley scattering between
$\boldsymbol{K}$ and $\boldsymbol{K'}$.

Decomposing the momentum sums into valley contributions,
$\sum_{\boldsymbol{k}} = \sum_{\boldsymbol{k}}^{\boldsymbol{K}} +
\sum_{\boldsymbol{k}}^{\boldsymbol{K'}}$, yields eight distinct terms
corresponding to intra- and intervalley scattering. Following the
continuum formulation derived in~\cite{abulafiya2026}, the full
Hamiltonian $\hat{H}=\hat{H}_0+\hat{V}_{\mathrm{vac}}$ takes, near the
Dirac points, the first-quantized form
\begin{equation}
  \hat{H}_0 =
    \begin{pmatrix}
       \Delta & 0 & \hat{L} & 0 \\
       0 & -\Delta & 0 & -\hat{L}^\dagger \\
       \hat{L}^\dagger & 0 & -\Delta & 0 \\
       0 & -\hat{L} & 0 & \Delta
    \end{pmatrix},
  \label{eq:hal_operator}
\end{equation}
with $\hat{L} = v_F(i\partial_x - \partial_y)$ and
$\Delta = 3\sqrt{3}\,t_2\sin\phi$, and the vacancy contribution
\begin{equation}
  \hat{V}_\mathrm{vac} =
    \begin{pmatrix}
      2\Delta\delta(\boldsymbol{r}) & 0
        & -\delta(\boldsymbol{r})\hat{L} & \delta(\boldsymbol{r})\hat{L}^\dagger \\
      0 & -2\Delta\delta(\boldsymbol{r})
        & \delta(\boldsymbol{r})\hat{L} & \delta(\boldsymbol{r})\hat{L}^\dagger \\
      -\hat{L}^\dagger\delta(\boldsymbol{r}) & -\hat{L}^\dagger\delta(\boldsymbol{r})
        & 0 & 0 \\
      \hat{L}\delta(\boldsymbol{r}) & \hat{L}\delta(\boldsymbol{r})
        & 0 & 0
    \end{pmatrix}.
  \label{eq:vac_operator}
\end{equation}

The role played by the Bloch Hamiltonian in a defect-free system generalizes to
the Weyl symbol (phase-space symbol) of the position-dependent Hamiltonian in
a system with defects~\cite{Goft2023}.  Computing the symbol of
$\hat{H} = \hat{H}_0 + \hat{V}_\mathrm{vac}$, and retaining only the terms
relevant to topology (see Appendix~\ref{app:symbol}), gives
\begin{equation}
  \mathcal{H}(\boldsymbol{k},\boldsymbol{r})
    = \begin{pmatrix}
        -\Delta & 0 & k_x - ik_y & h(r)\,e^{i\theta} \\
        0 & \Delta & h(r)\,e^{-i\theta} & -k_x - ik_y \\
        k_x + ik_y & h(r)\,e^{i\theta} & \Delta & 0 \\
        h(r)\,e^{-i\theta} & -k_x + ik_y & 0 & -\Delta
      \end{pmatrix}
  \label{eq:symbol}
\end{equation}
where $\psi(\boldsymbol{r}) = h(r)\,e^{i\theta}$ is a localized complex
scalar field describing the vacancy, with $h(r) \geq 0$ its real-valued
radial profile.  The angular factor $e^{i\theta}$ encodes the nontrivial
phase winding of the defect.

For $N_A$ vacancies on the $A$-sublattice and $N_B$ on the $B$-sublattice,
the defect field generalizes to
\begin{equation}
  \psi(\boldsymbol{r}) = h(r)\, e^{i\theta(N_A - N_B)},
  \label{eq:psi_multi}
\end{equation}
with the total phase winding number $m = N_A - N_B$.

\subsection{The \texorpdfstring{$\mathbb{Z}_2$}{Z2} invariant}

The topological invariant associated with the vacancy configuration is
obtained by expressing the angular dependence of \eqref{eq:psi_multi} as a
gauge transformation acting on the clean Hamiltonian.  This transformation is
not single-valued and hence encodes nontrivial topology.  Evaluating the
Chern-Simons formula on the doubled Hilbert space (see Appendix~\ref{app:z2})
yields the central result of this paper,
\begin{equation}
  \nu = C\cdot m \;\; \mathrm{mod}\; 2 = |N_A - N_B|\;\;\mathrm{mod}\;2,
  \label{eq:z2}
\end{equation}
where $C = \pm 1$ is the Chern number of the bulk Haldane model.

According to the Atiyah-Singer index
theorem~\cite{Atiyah1963,Atiyah1968,Freed2021,Eguchi1980,Roe1990,Nakahara1990},
$\nu$ equals the number of zero-energy modes modulo 2.  Consequently:
\begin{itemize}[nosep]
  \item $\nu=1$ (odd $|N_A-N_B|$): one protected zero-energy mode exists.
  \item $\nu=0$ (even $|N_A-N_B|$): no protected zero mode; vacancy-induced
    states hybridize and split away from zero energy.
\end{itemize}
This even-odd effect is a direct consequence of particle-hole symmetry, which
enforces a spectrum symmetric about $E=0$.  An odd number of defect-induced
states cannot all be paired, and one must remain pinned at zero energy.

\section{Spectral Properties and Zero Modes}
\label{sec:modes}

\begin{figure}[b]
\centering
\includegraphics[width=0.53\linewidth]{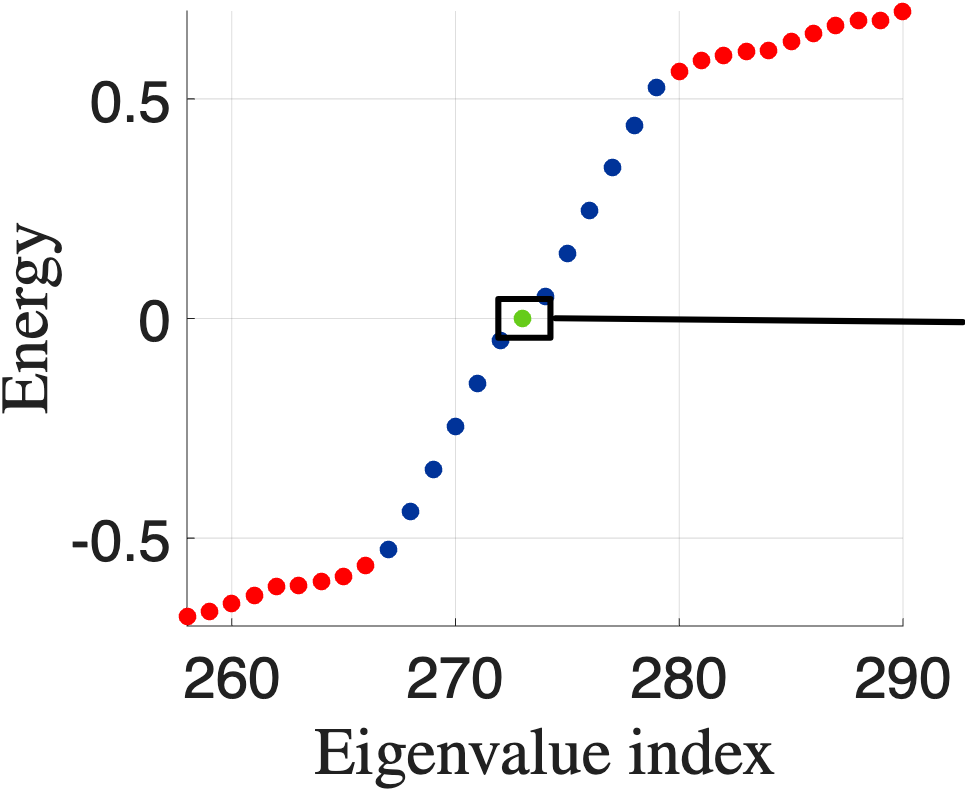}
\hfill
\includegraphics[width=0.42\linewidth]{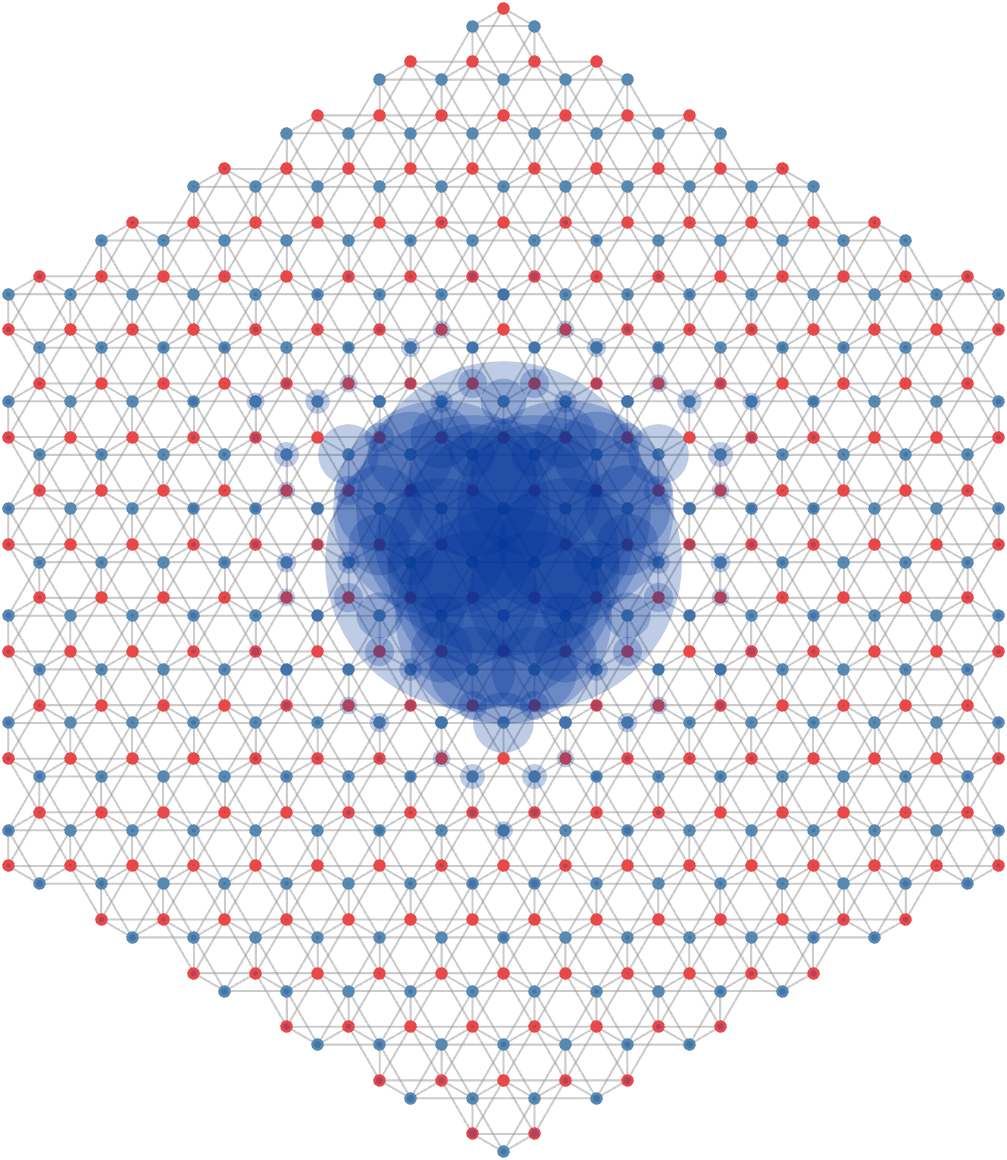}
\caption{Haldane model with a single vacancy ($\nu=1$).  Left: energy
  spectrum showing a zero-energy defect mode (green), chiral edge states
  (blue), and bulk modes (red).  Right: spatial probability density
  $|\psi|^2$ of the zero mode, demonstrating localization at the vacancy
  site.  The presence of bulk-topology edge states confirms that the bulk gap
  and Chern number are unaffected by the vacancy.}
\label{fig:1vac}
\end{figure}

Figure~\ref{fig:1vac} shows the spectrum and wavefunction of a system with
a single vacancy ($N_A=1$, $N_B=0$, $m=1$, $\nu=1$).  A state is pinned at
exactly zero energy, in addition to the chiral edge states that remain
present inside the bulk gap.  The defect-induced zero mode is spatially
localized at the vacancy site, whereas the edge states retain their
boundary-localized character.  The coexistence of both types of states
confirms that the vacancy does not destroy the underlying bulk topology but
instead introduces an additional in-gap mode.

\begin{figure}[t]
\centering
\includegraphics[width=0.53\linewidth]{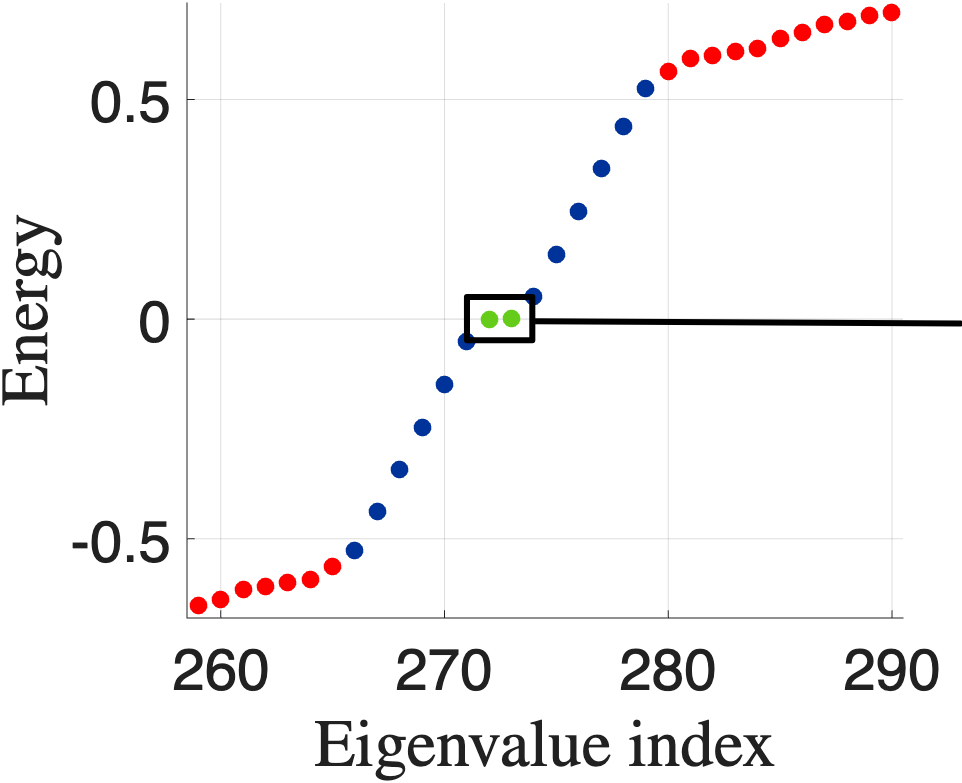}
\hfill
\includegraphics[width=0.42\linewidth]{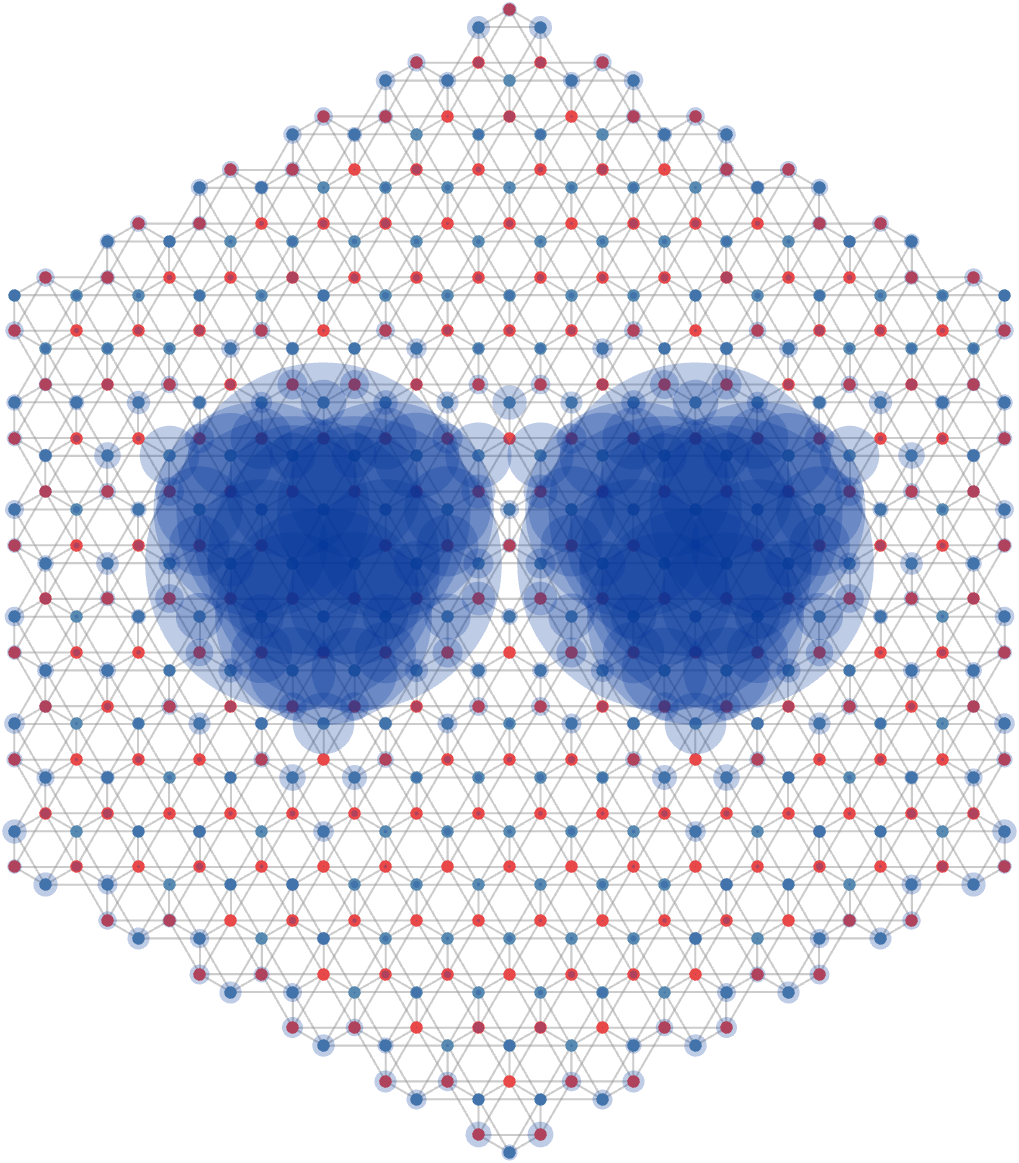}
\caption{Haldane model with two vacancies on the same sublattice ($N_A=2$,
  $N_B=0$, $\nu=0$).  Left: energy spectrum showing two defect-induced modes
  (green) that hybridize and split symmetrically about zero energy, with edge
  states (blue) and bulk modes (red) also visible.  The two modes are lifted
  from zero energy because the $\mathbb{Z}_2$ invariant $\nu=0$ provides no
  topological protection.  Right: spatial probability density of one of the
  hybridized modes, showing localization around the two vacancy sites.}
\label{fig:2vac}
\end{figure}

Figure~\ref{fig:2vac} shows the two-vacancy case with both vacancies on the
same sublattice ($N_A=2$, $N_B=0$, $m=2$, $\nu=0$).  The spectrum now
exhibits two in-gap modes that hybridize and split symmetrically about zero
energy: no state is pinned at $E=0$.  The corresponding wavefunctions remain
localized at the vacancy sites, confirming their defect origin, but the
$\mathbb{Z}_2$ invariant $\nu=0$ correctly predicts the absence of topological
protection.  The same qualitative behavior is observed for two vacancies on opposite sublattices ($N_A=N_B=1$, $m=0$, $\nu=0$): the two modes hybridize
and no zero-energy state survives.

\section{Wavefunction Dislocations}
\label{sec:dislocations}

A key diagnostic of nontrivial defect topology is the presence of
\emph{dislocations} in the wavefunction at the defect
site~\cite{abulafia_wavefronts_2023}.  The underlying idea is that
interference between the two inequivalent valleys $\boldsymbol{K}$ and
$\boldsymbol{K'}$ of the Brillouin zone produces oscillatory wavefronts with
wavelength $\lambda = 2\pi/|\boldsymbol{K}-\boldsymbol{K'}|$.  When a defect
carries a nonzero phase winding, it forces a mismatch in these wavefronts that
appears as a dislocation in the real-space wavefunction.

To make the dislocations visible, we proceed as follows.  We obtain the
eigenstate $\psi(\boldsymbol{R}_n)$ on the discrete lattice sites by
numerical diagonalization, interpolate to a smooth real-space function
$\psi(\boldsymbol{r})$, and Fourier transform.  We then filter the Fourier
transform around both valleys $\boldsymbol{K}$ and $\boldsymbol{K'}$ and perform an
inverse Fourier transform to obtain a filtered wavefunction that isolates the
intervalley interference pattern.

\begin{figure}[t]
\centering
\begin{minipage}[t]{0.32\linewidth}\centering
  \includegraphics[height=2.8cm]{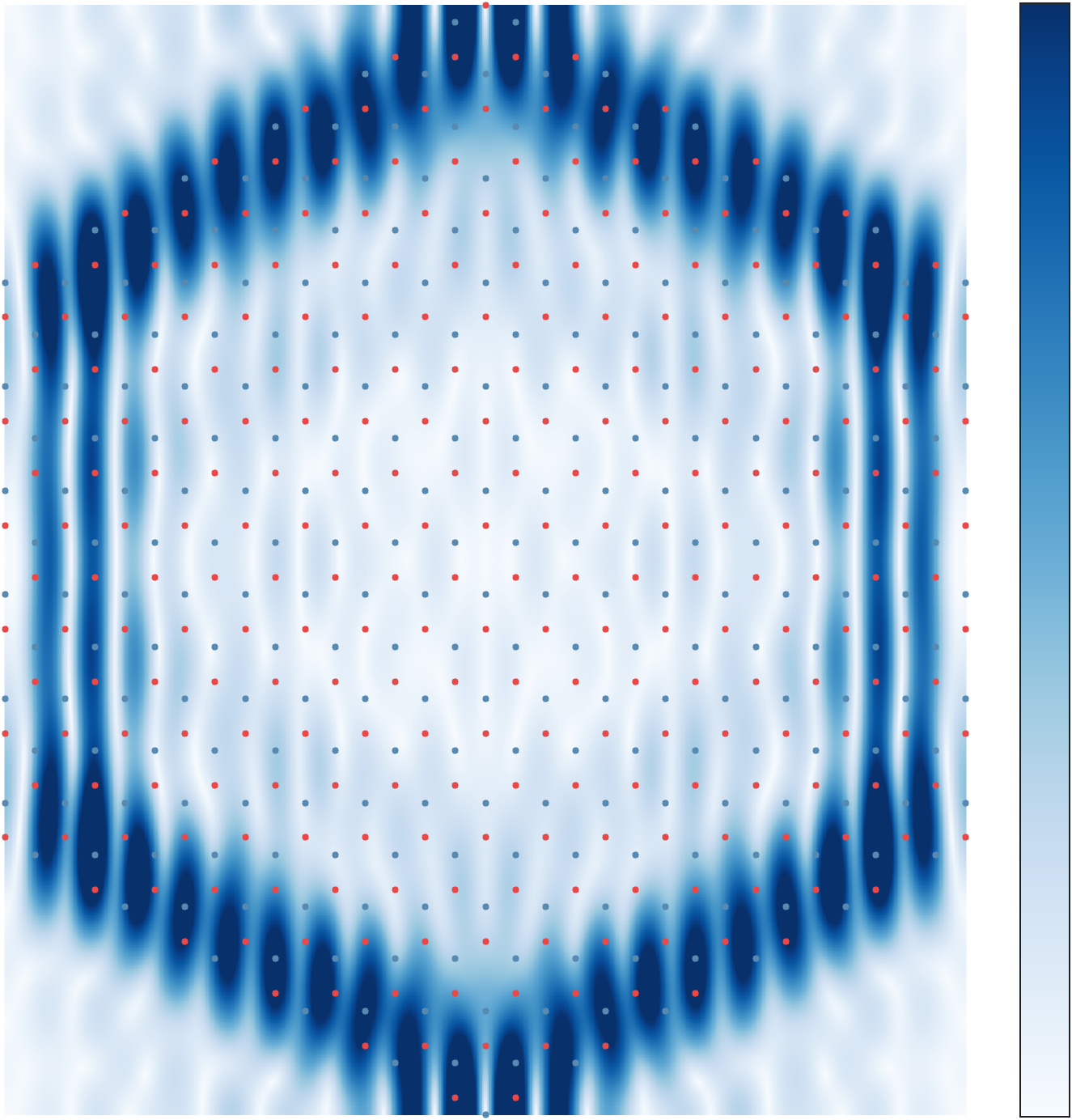}\\
  \hspace*{-3mm}\small\text{(a)}
\end{minipage}\hfill
\begin{minipage}[t]{0.32\linewidth}\centering
  \includegraphics[height=2.8cm]{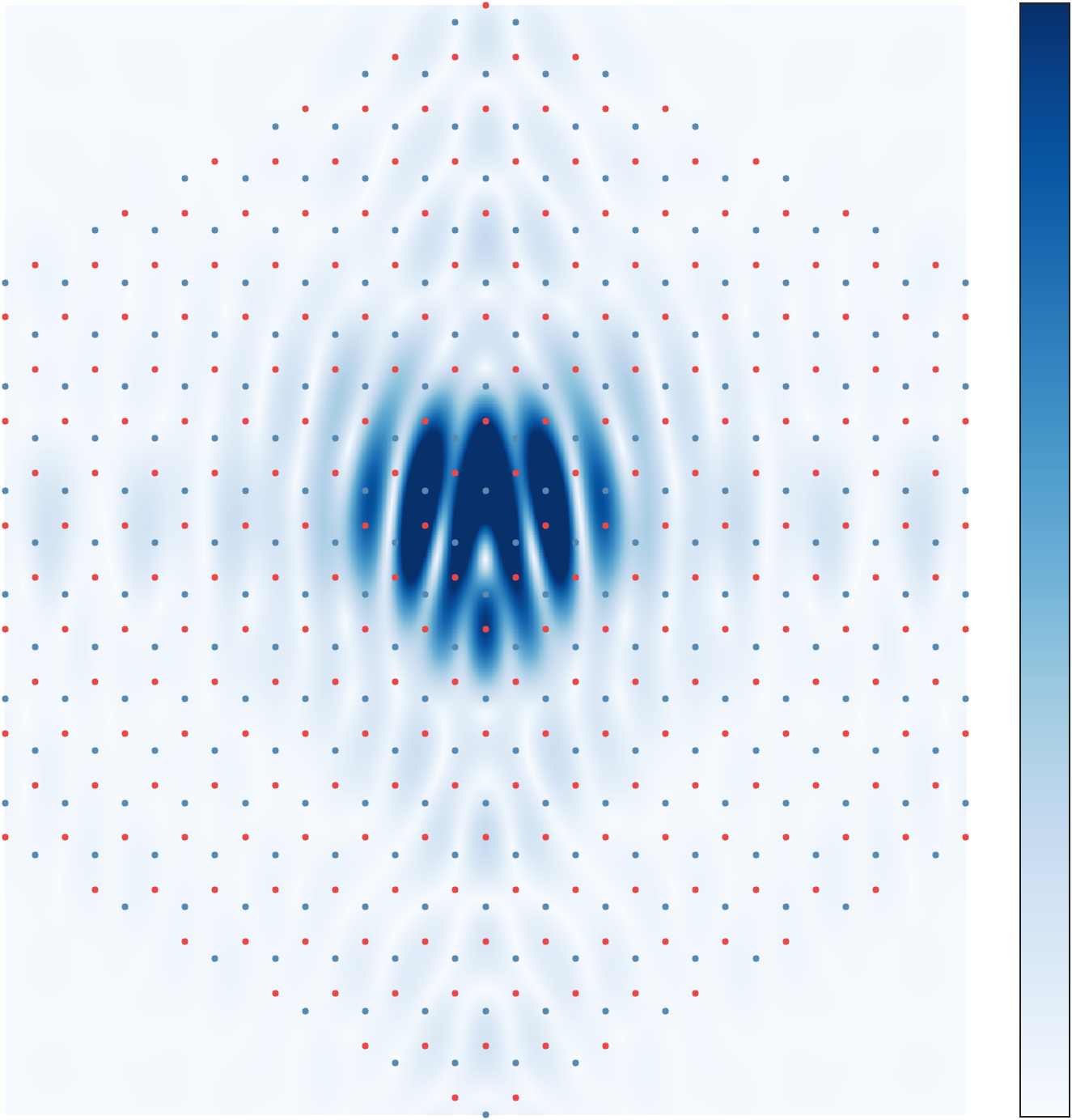}\\
  \hspace*{-3mm}\small\text{(b)}
\end{minipage}\hfill
\begin{minipage}[t]{0.32\linewidth}\centering
  \includegraphics[height=2.8cm]{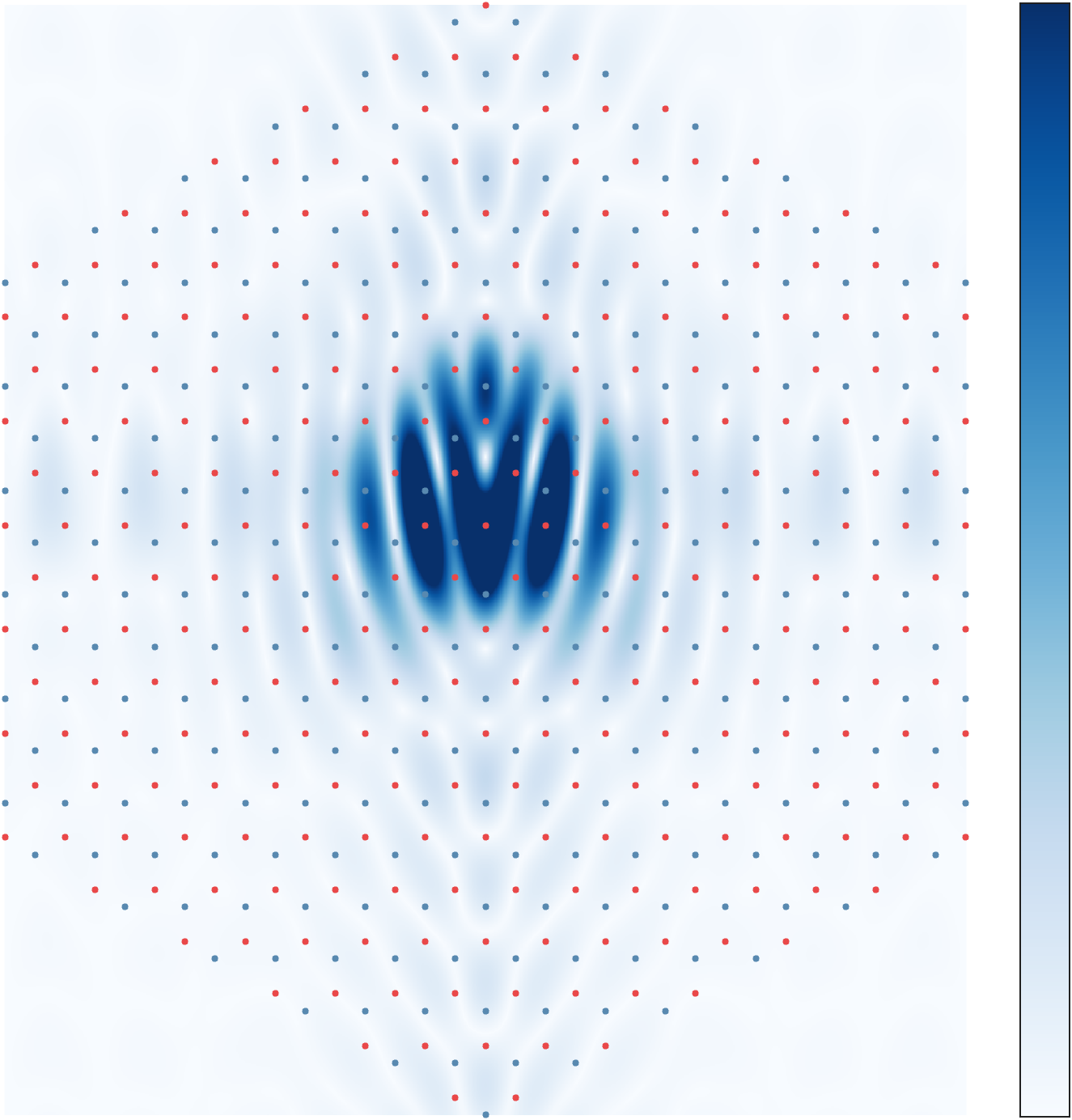}\\
  \hspace*{-3mm}\small\text{(c)}
\end{minipage}

\begin{minipage}[t]{0.32\linewidth}\centering
  \includegraphics[height=2.8cm]{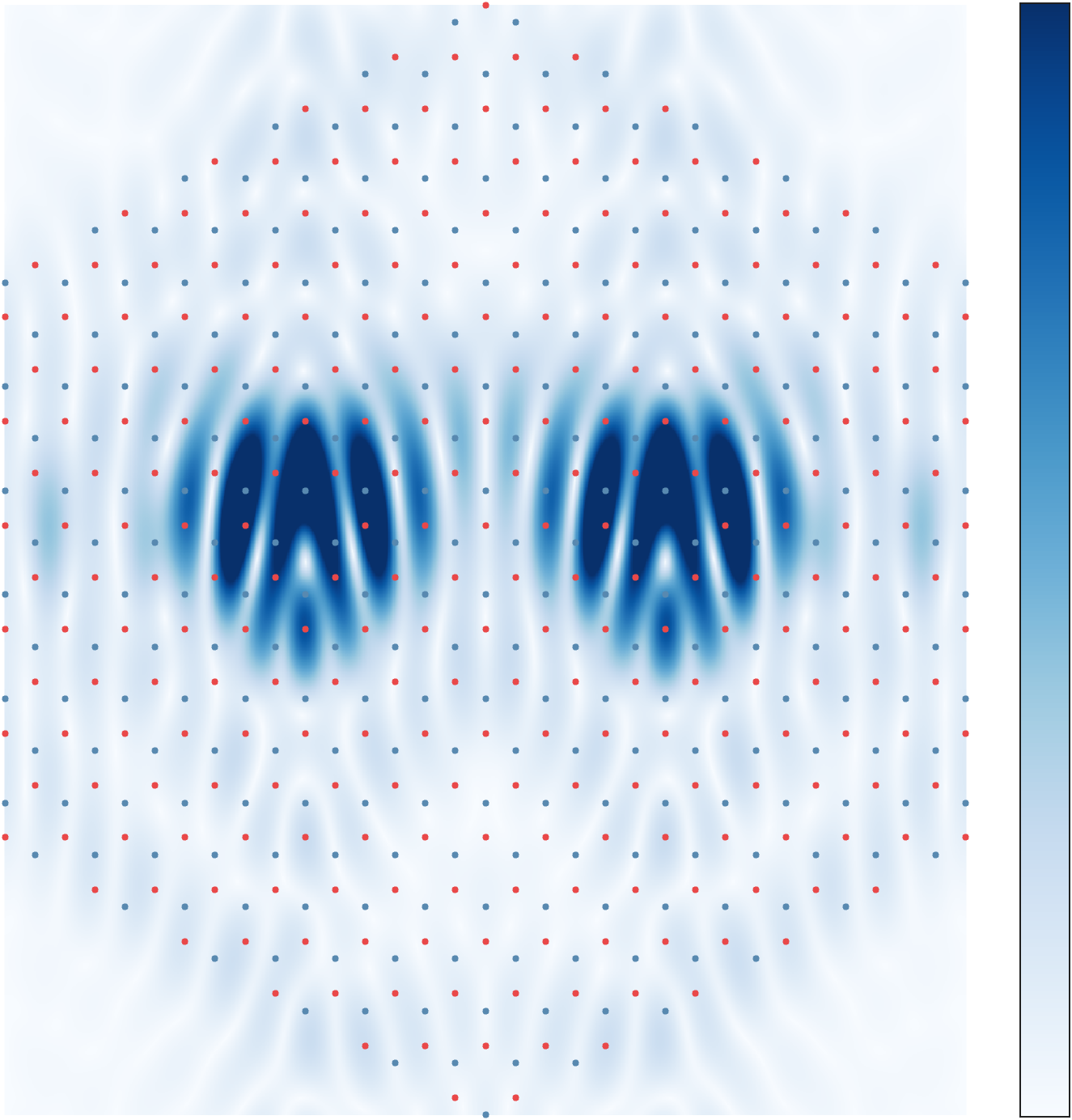}\\
  \hspace*{-3mm}\small\text{(d)}
\end{minipage}\hfill
\begin{minipage}[t]{0.32\linewidth}\centering
  \includegraphics[height=2.8cm]{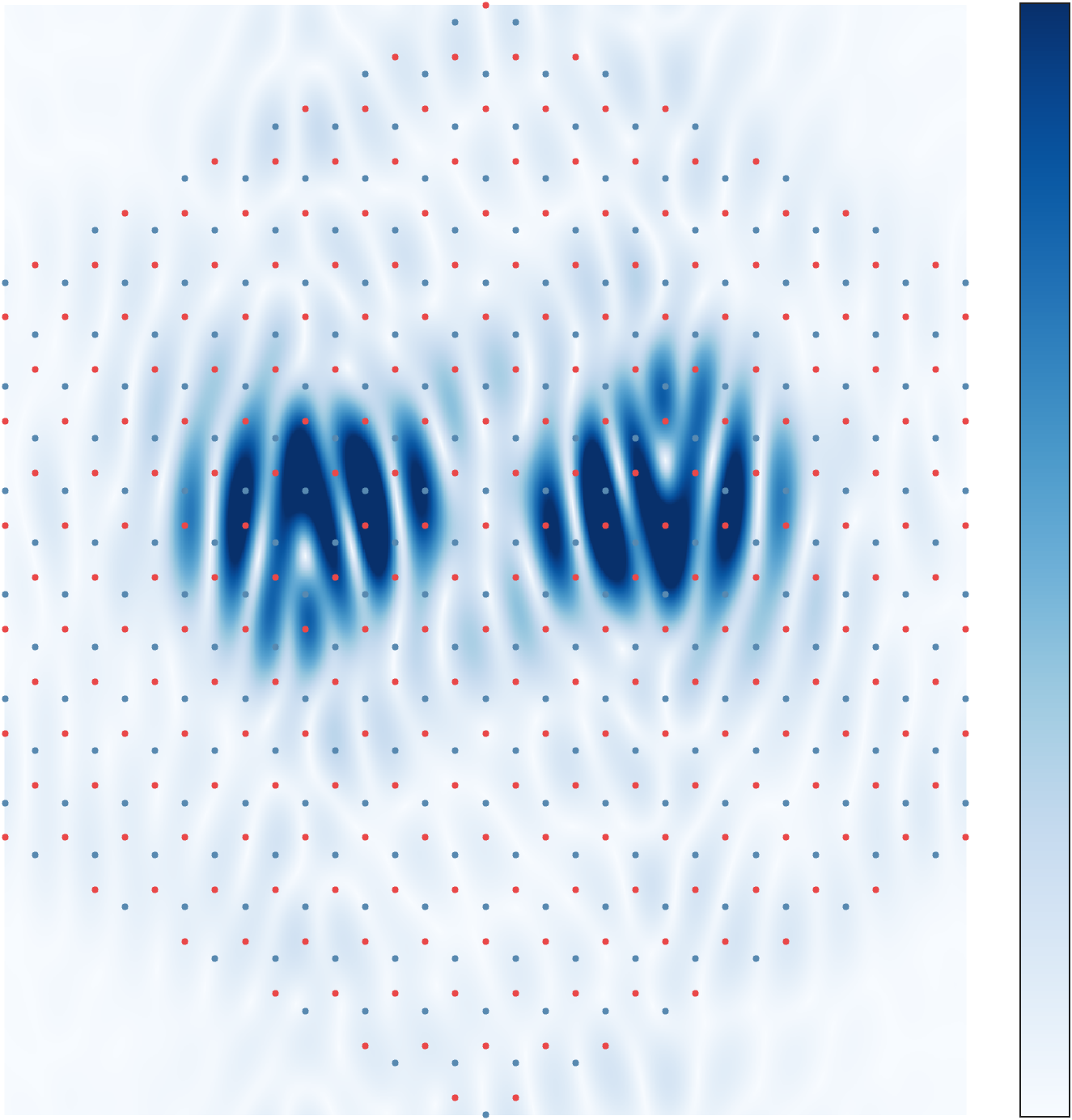}\\
  \hspace*{-3mm}\small\text{(e)}
\end{minipage}\hfill
\begin{minipage}[t]{0.32\linewidth}\centering
  \includegraphics[height=2.8cm]{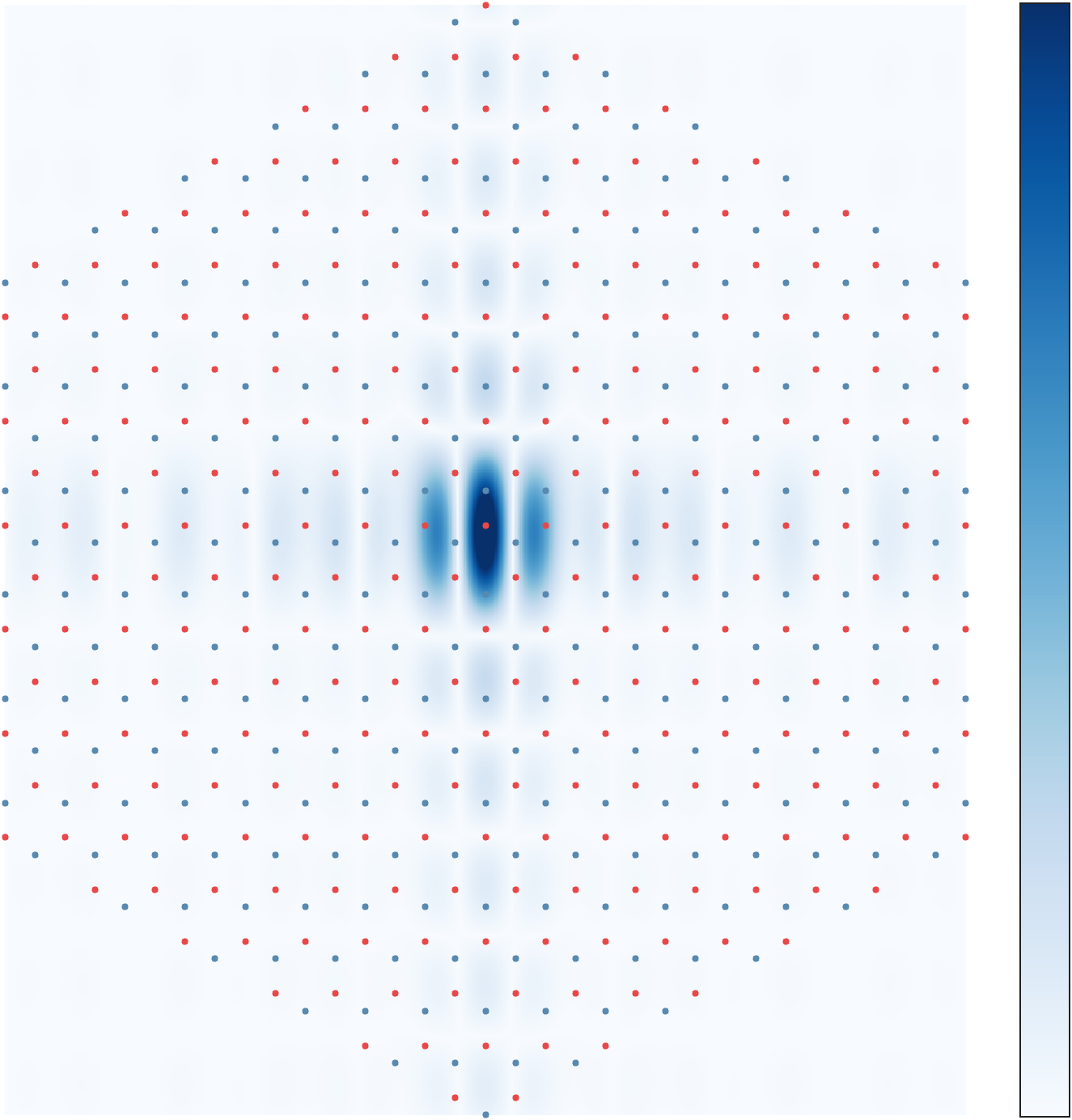}\\
  \hspace*{-3mm}\small\text{(f)}
\end{minipage}
\caption{Filtered wavefronts for six defect configurations, illustrating
the $\mathbb{Z}_2$ pattern. (a) Edge state, no defect: smooth wavefronts,
no dislocation. (b) Single $A$-sublattice vacancy ($\nu=1$): a single
dislocation is present. (c) Single $B$-sublattice vacancy ($\nu=1$): a
single dislocation is present. (d) Two $A$-sublattice vacancies
($\nu=0$): two dislocations are visible. (e) One $A$- and one
$B$-sublattice vacancy ($\nu=0$): two dislocations are visible.
(f) Adatom (trivial defect): no dislocation. Since the topological
classification is $\mathbb{Z}_2$, dislocations are counted modulo two;
thus configurations with an even number of dislocations belong to the
trivial sector ($\nu=0$), whereas configurations with an odd number of
dislocations belong to the nontrivial sector ($\nu=1$). The observed
wavefront patterns therefore agree with Eq.~\eqref{eq:z2}.}

\label{fig:dislocations}
\end{figure}

Figure~\ref{fig:dislocations} shows the results for six distinct
configurations, arranged to illustrate the $\mathbb{Z}_2$ pattern.
The reference edge state (no vacancy, panel (a)) and adatom defect (panel (f))
show smooth, uninterrupted wavefronts with no dislocation. Single vacancies
on the $A$- or $B$-sublattice (panels (b) and (c)) each yield
$\nu=1$ and display a clear dislocation at the defect site, consistent with
a phase winding $m=\pm 1$. In contrast, two vacancies on the same sublattice (panel (d)) and one vacancy on each sublattice (panel (e)) both correspond to $\nu=0$ and exhibit an even number of dislocations. Since the
topological classification is $\mathbb{Z}_2$, dislocations are counted only
modulo two: pairs of dislocations are topologically equivalent to the absence
of a dislocation and therefore correspond to the trivial sector. Thus, while
individual dislocations remain visible in the wavefront pattern, their net
topological charge vanishes, consistent with $\nu=0$.

This systematic pattern -- an odd number of dislocations if and only if
$\nu=1$ -- provides a direct real-space visualization of the $\mathbb{Z}_2$
invariant. Crucially, the adatom in panel (f) also shows no dislocation,
confirming that the effect is not generic to any local perturbation but is
specific to topological defects.

\section{Fractional Charge}
\label{sec:charge}

A complementary signature of the topological zero mode is the accumulation
of a fractional charge at the vacancy
site~\cite{Jackiw1976,Hou2007,Seradjeh2008,Obispo2015,Lee2020,Peterson2021}.
The induced charge is defined as the difference between the accumulated charge densities of the systems with and without the defect~\cite{ovdat2020},
\begin{equation}
  Q = \sum_{E_n \leq 0}\int_{\mathcal{D}} d^2r\,
      \Big(|\psi_n(\boldsymbol{R})|^2 - |\psi_n^{(0)}(\boldsymbol{R})|^2\Big)
    = -\frac{e}{2}\,\nu,
  \label{eq:frac_charge}
\end{equation}
where the sum runs over all occupied states ($E_n\leq 0$), $\mathcal{D}$ is a
disk centered at the defect, and $\nu$ is the $\mathbb{Z}_2$ invariant.  For
topological defects ($\nu=1$), the induced charge is $-e/2$; for trivial
defects ($\nu=0$), no net charge accumulates.

\begin{figure}[t]
\centering
\includegraphics[width=0.49\linewidth]{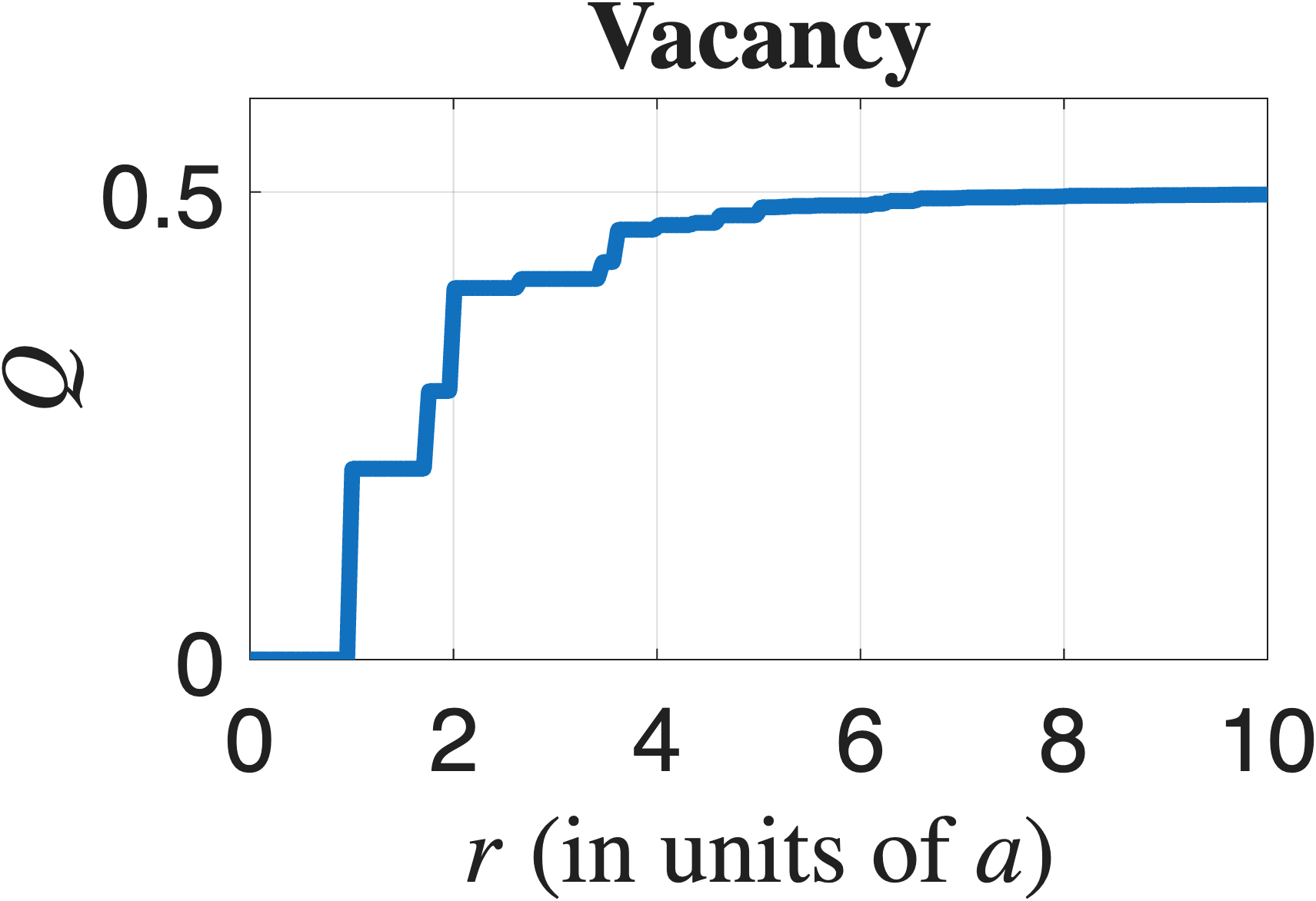}
\hfill
\includegraphics[width=0.49\linewidth]{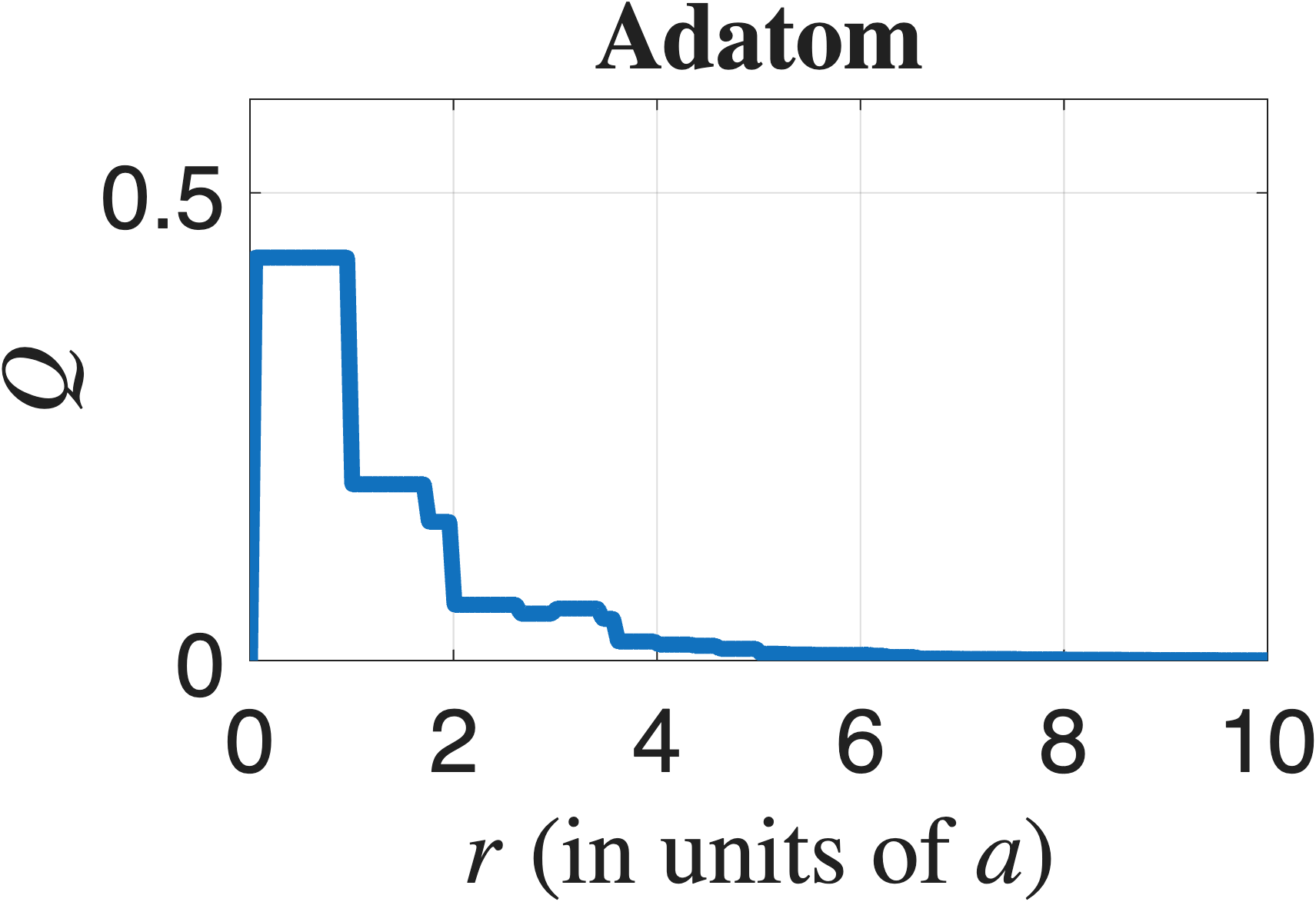}
\caption{Induced charge $Q$ as a function of the radius of the integration
  disk $\mathcal{D}$, in units of the lattice constant $a$.  Left: single
  vacancy ($\nu=1$).  The accumulated charge saturates to $-e/2$ at large
  radii, confirming the topological prediction of Eq.~\eqref{eq:frac_charge}.
  Right: adatom ($\nu=0$).  No net charge accumulates, consistent with the
  topologically trivial nature of the adatom defect.}
\label{fig:frac_charge}
\end{figure}

Figure~\ref{fig:frac_charge} shows the integrated induced charge as a function
of disk radius.  For the vacancy, $Q$ saturates to $-e/2$, in quantitative
agreement with Eq.~\eqref{eq:frac_charge}.  For the adatom, $Q$ decreases to zero and shows no net accumulation, confirming the trivial character of
the adatom-induced state.  This provides an independent, experimentally
measurable criterion to distinguish topological from trivial defect-induced
states, complementary to the wavefunction dislocation analysis of
Sec.~\ref{sec:dislocations}.

\section{Chiral Currents}
\label{sec:currents}

The final signature we investigate is the probability-current associated with
defect-induced and edge states.  We define the bond current from site $i$ to
site $j$ as
\begin{equation}
  J_{i\to j} = 2\,\Im\!\left(\psi_i^*\, t_{ij}\, \psi_j\right),
  \label{eq:current}
\end{equation}
where $t_{ij}$ is the hopping amplitude and $\psi$ is the relevant eigenstate.

\begin{figure}[t]
\centering
\begin{minipage}[t]{0.32\linewidth}
\centering
\includegraphics[width=\linewidth,height=2.8cm]{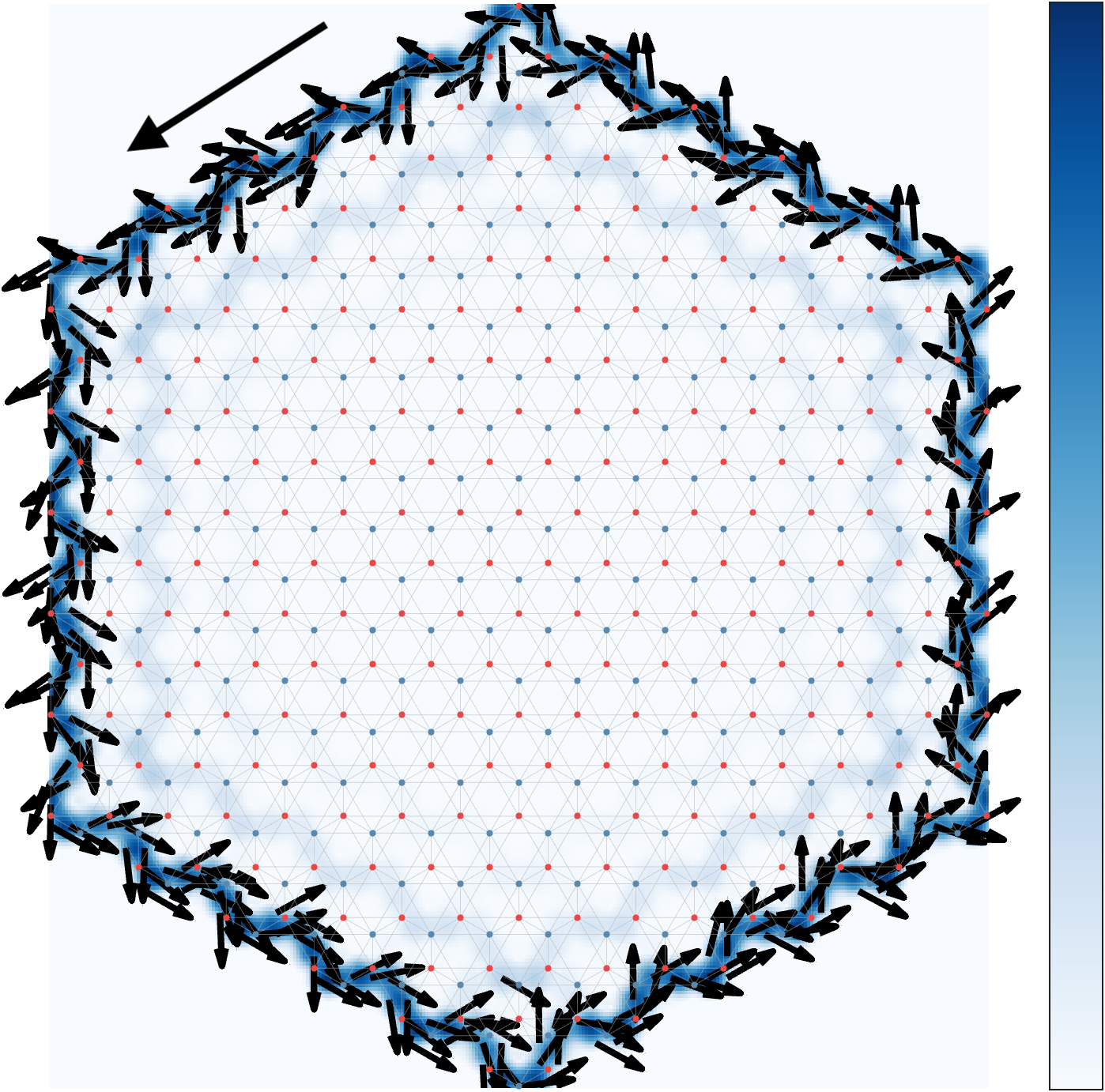}\\
\hspace*{-2mm}\small\text{(a)}
\end{minipage}\hfill
\begin{minipage}[t]{0.32\linewidth}
\centering
\includegraphics[width=\linewidth,height=2.8cm]{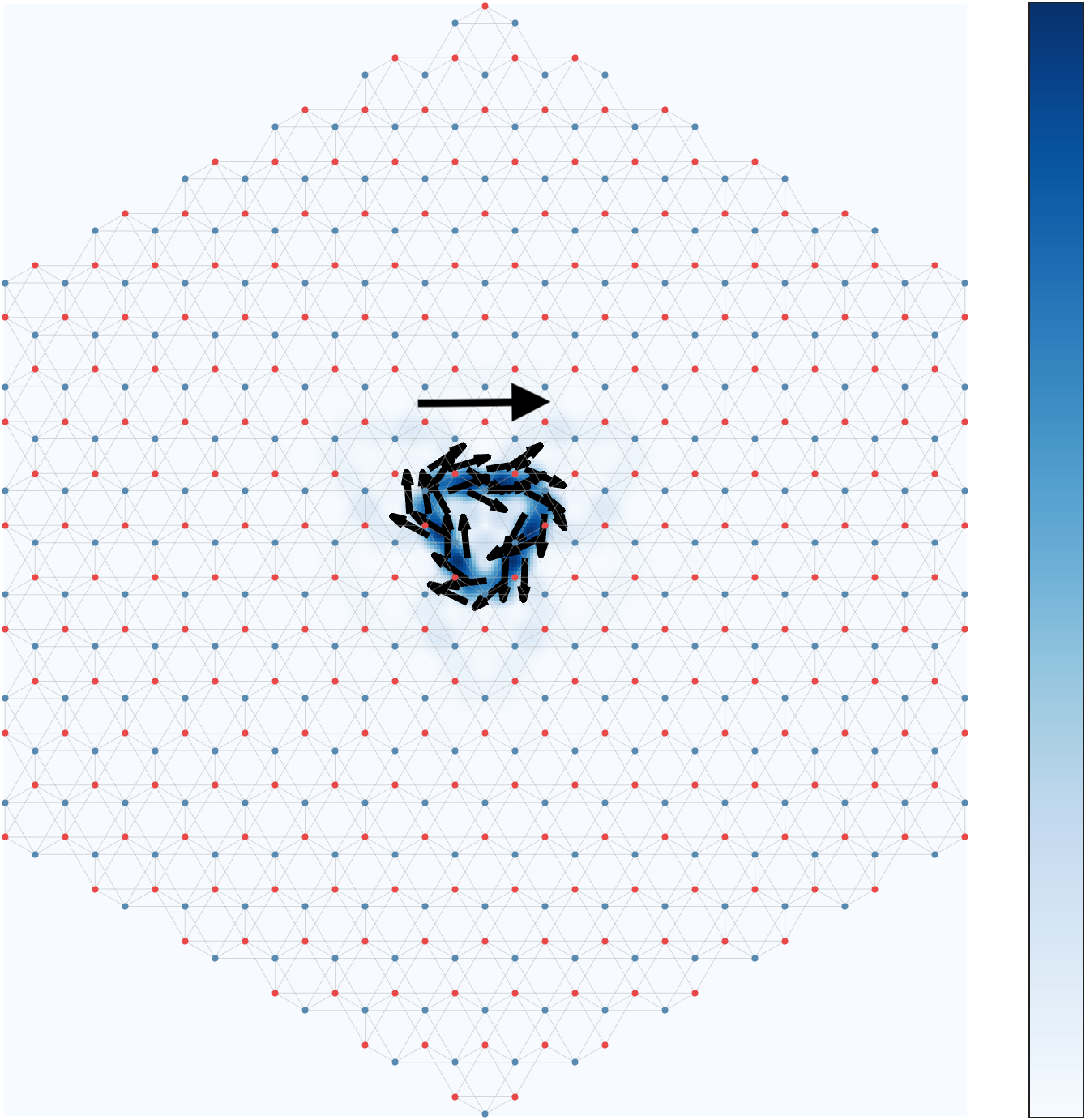}\\
\hspace*{-3mm}\small\text{(b)}
\end{minipage}\hfill
\begin{minipage}[t]{0.32\linewidth}
\centering
\includegraphics[width=\linewidth,height=2.8cm]{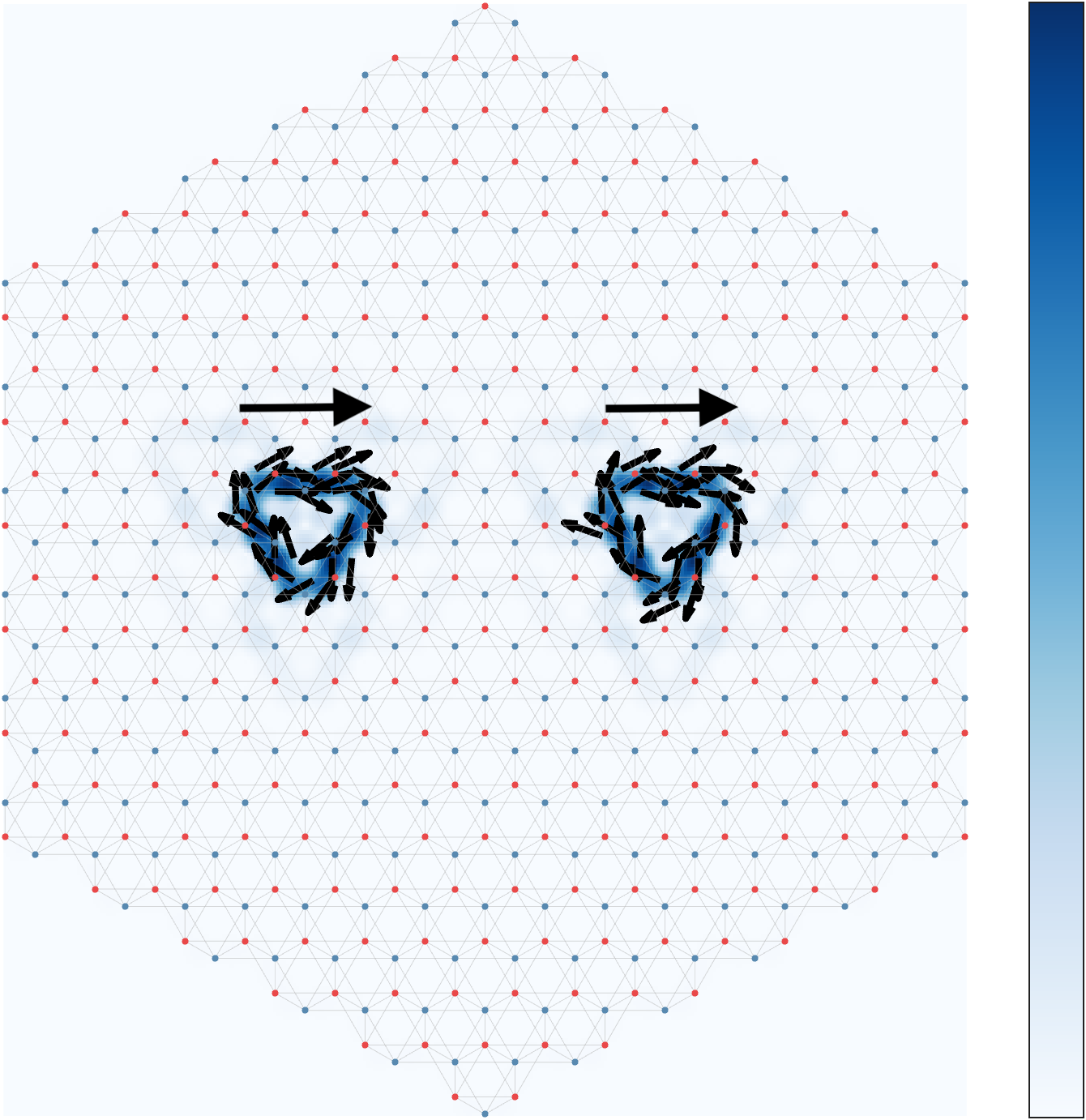}\\
\hspace*{-3mm}\small\text{(c)}
\end{minipage}
\caption{Probability current patterns computed from Eq.~\eqref{eq:current}
  for three representative states.
  (a) Edge state of the clean Haldane model ($C=+1$): counterclockwise
  (positive) circulation consistent with the chiral edge transport.
  (b) Zero mode of a single-vacancy system ($\nu=1$): clockwise (negative)
  circulation, \emph{opposite} to the edge current.
  (c) One of the hybridized modes of a two-vacancy system ($\nu=0$): current
  circulation also opposite to the edge, even though the mode is no longer
  at zero energy.
  The current reversal in panels (b) and (c) directly reflects the
  topological charge of the vacancy and is the tight-binding analogue of
  the current reversal produced by a vortex in a $p$-wave superconductor.}
\label{fig:currents}
\end{figure}

Figure~\ref{fig:currents} compares the current patterns for three states:
an edge state of the clean system, the zero mode of a single vacancy, and one
of the hybridized modes of a two-vacancy system.

In the clean Haldane model at $\phi=\pi/2$ ($C=+1$), the chiral edge state
circulates counterclockwise [Fig.~\ref{fig:currents}(a)], consistent with
the bulk-boundary correspondence.  Changing to $\phi=-\pi/2$ ($C=-1$)
reverses the edge circulation, as expected.

The single-vacancy zero mode [Fig.~\ref{fig:currents}(b)] produces a current
that circulates \emph{clockwise}, i.e., in the direction \emph{opposite} to the
edge current.  This reversal is not an artifact of the geometry but a
topological effect: the vacancy carries a $\mathbb{Z}_2$ charge $\nu=1$ that
forces the defect-induced state to circulate oppositely to states arising from
the bulk topology.  Changing $\phi\to -\phi$ reverses both currents
simultaneously, confirming that their relative orientation is locked by
topology and not by any particular parameter choice.

The two-vacancy case [Fig.~\ref{fig:currents}(c)] shows that current reversal
persists even when $\nu=0$: the hybridized modes are no longer at zero energy,
but they still retain the opposite circulation inherited from the topological
character of the individual vacancy states before hybridization.

\section{Discussion}
\label{sec:discussion}

\paragraph{Three independent signatures.}
The results of Secs.~\ref{sec:modes}--\ref{sec:currents} establish that three
independent diagnostics -- wavefunction dislocations,
fractional charge, and current reversal -- all agree on the same
$\mathbb{Z}_2$ classification.  Each diagnostic provides a distinct physical
window onto the same underlying topological invariant $\nu = C\cdot m\,
\mathrm{mod}\,2$, and each sharply discriminates vacancies ($\nu=1$ for odd
$|m|$) from adatoms ($\nu=0$ always).

\paragraph{Analogy with \boldmath$p$-wave superconductors.}
The current reversal in Sec.~\ref{sec:currents} is precisely analogous to a
well-known phenomenon in topological superconductivity.  A spinless $p+ip$
superconductor in two dimensions (symmetry class D) has Chern number $C=1$
and binds a Majorana zero mode to each vortex~\cite{gurarie2007}.  The vortex
carries a $\mathbb{Z}_2$ invariant; the Majorana mode produces a half-integer
charge $e/2$ at the vortex core and its associated current opposes the
supercurrent of the condensate.  In the present system, the vacancy plays the
role of the vortex, the chiral edge current plays the role of the condensate
supercurrent, and the $\mathbb{Z}_2$ invariant $\nu$ determines whether such
a mode exists.  The Haldane model differs from the superconductor in that
charge is conserved (no Majorana fermions), but the topological structure is
otherwise identical.  This analogy places our results within a broader
framework: the bulk-defect coupling formula \eqref{eq:z2} is a general
consequence of the tenfold classification and should apply to any system where
a class-D bulk $\mathbb{Z}$ invariant coexists with a class-D defect
$\mathbb{Z}_2$ invariant.

\paragraph{Even-odd effect and sublattice control.}
Equation~\eqref{eq:z2} implies a direct experimental handle on defect topology:
placing vacancies on the same versus opposite sublattices controls $\nu$.
Specifically, all vacancies on one sublattice give $|m|=N$, and $\nu=N\,
\mathrm{mod}\,2$.  Equal number of vacancies on each sublattice gives $m=0$, $\nu=0$,
regardless of the number of vacancies.  This sublattice selectivity provides a
means to switch defect topology on and off by design~\cite{ovdat2020}.

\paragraph{Experimental platforms.}
The Haldane model has been realized in several controllable platforms.  In
ultracold fermionic atoms, the model was first implemented using periodically
modulated optical lattices~\cite{jotzu2014}, with precise control over both
hopping parameters and the phase $\phi$.  Photonic implementations using
arrays of coupled resonators with broken time-reversal symmetry provide another
clean realization. In both settings, single-site vacancies can be engineered
with high precision, and the three signatures identified here are in principle
directly measurable: dislocations by variations of local electronic density~\cite{abulafia_wavefronts_2023}, fractional charge by
integrated local density of states, and current patterns by measuring
time-of-flight or local flow observables.  In electronic systems on metal
surfaces, engineered honeycomb lattices of adsorbed atoms offer yet another
platform where individual-site control has been demonstrated.

\section{Conclusion}
\label{sec:conclusion}

We have studied the coexistence of bulk and defect topology in the Haldane
model.  The bulk system has Chern number $C=\pm 1$; point vacancies introduce
an additional $\mathbb{Z}_2$ topological structure characterized by the
invariant $\nu = C\cdot m\,\mathrm{mod}\,2$, where $m=N_A-N_B$ is the net
sublattice imbalance.  This invariant controls, via the Atiyah-Singer index
theorem, the existence of protected zero-energy modes.

We identified and verified three complementary signatures of the defect
topology, each distinguishing vacancies (topological) from adatoms (trivial):
(i) wavefunction dislocations that track the phase winding of the defect
configuration; (ii) fractional charge $e/2$ localized at vacancy sites; and
(iii) probability currents that circulate opposite to chiral edge states, in
direct analogy with the current reversal produced by a vortex in a $p$-wave
superconductor.

All three signatures are captured by a single formula, Eq.~\eqref{eq:z2},
which encodes the coupling between the bulk $\mathbb{Z}$ invariant and the
defect $\mathbb{Z}_2$ invariant.  We anticipate that this bulk-defect coupling
principle extends beyond the Haldane model to any system in the tenfold
classification where a bulk $\mathbb{Z}$ invariant coexists with a defect
$\mathbb{Z}_2$ invariant.

\section*{Acknowledgments}
This work was supported by the Israel Science Foundation under Grant No. 772/21 and by the Pazy Foundation. It is a pleasure to thank Y. Abulafia and S. Faktor for helpful discussions.
\appendix

\section{Symbol (Weyl transform) of the Vacancy Hamiltonian}
\label{app:symbol}

In this appendix, we derive the symbol \eqref{eq:symbol} used in
Sec.~\ref{sec:defects}. The Weyl symbol provides a phase-space representation of the
Hamiltonian and serves as the starting point for the extraction of
topological invariants from spatially inhomogeneous systems. This
approach was introduced in~\cite{Goft2023}. The Weyl symbol of a position-dependent Hamiltonian $\hat{H}$ is defined by
\begin{equation}
  \mathcal{H}(\boldsymbol{k},\boldsymbol{r})
    = \int d^2\xi\; e^{i\boldsymbol{k}\cdot\boldsymbol{\xi}}\,
      \Bigl\langle \boldsymbol{r}-\tfrac{\boldsymbol{\xi}}{2}\Big|
        \hat{H}
      \Big|\boldsymbol{r}+\tfrac{\boldsymbol{\xi}}{2}\Bigr\rangle.
\end{equation}
For $\hat{H}_0$ the symbol reproduces the Bloch Hamiltonian.  The vacancy
contribution \eqref{eq:vac_operator} involves $\delta(\boldsymbol{r})$, so its
symbol acquires position dependence.

In the low-energy (continuum Dirac) limit, and working in the valley-doubled
basis $(\psi_K, \psi_{K'})$, the clean Hamiltonian near the Dirac points can
be written in the form $\mathcal{H}_0 = k_x\sigma_x\otimes\sigma_z +
k_y\sigma_y\otimes\mathbf{1} - \Delta\sigma_z\otimes\sigma_z$.  The vacancy
potential, expressed as a spatially localized complex scalar field
$\psi(\boldsymbol{r}) = h(r)\,e^{i\theta}$, where $h(r) \geq 0$ is real-valued, acts as
\begin{equation}
  \mathcal{H} = \mathcal{H}_0 + h\,\sigma_x\otimes\sigma_x,
\end{equation}
and a unitary rotation $U(\theta) = \mathbf{1}\otimes e^{-i\theta\sigma_z/2}$
applied to the clean Hamiltonian reproduces this form when $U(\theta)$ winds
around the defect.  Since $U(\theta)$ is not single-valued for $\theta\in
[0,2\pi)$, the winding encodes nontrivial topology.  Using the identity
$e^{-i\theta\sigma_z/2}\sigma_x e^{i\theta\sigma_z/2} = \sigma_x\cos\theta
+ \sigma_y\sin\theta$ one recovers the symbol \eqref{eq:symbol} after
combining the contribution from both valleys, retaining only the topologically
relevant terms.

\section{Calculation of the \texorpdfstring{$\mathbb{Z}_2$}{Z2} Invariant}
\label{app:z2}

In this appendix, we derive Eq.~\eqref{eq:z2}.  Consider a class-D Hamiltonian
$H_p^0(\boldsymbol{k})$ in 2D with Chern number $p$.  We introduce a vacancy
by the gauge transformation
\begin{equation}
  H_p(\boldsymbol{k},\phi) = e^{-i\phi\tau_z/2}\, H_p^0(\boldsymbol{k})\,
                              e^{i\phi\tau_z/2},
\end{equation}
where $\phi\in[0,2\pi)$ parametrizes the phase winding around the defect and
$\tau_z$ acts in sublattice space.  The $\mathbb{Z}_2$ invariant is given by
the Chern-Simons formula,
\begin{equation}
  \nu = \frac{2}{d!}\!\left(\frac{i}{2\pi}\right)^{\!d}
        \int_{BZ^d\times S^{d-1}} Q_{2d-1}\;\;\mathrm{mod}\;2,
\end{equation}
evaluated on the base space $T^2\times S^1$ (2D Brillouin zone times the
loop parametrized by $\phi$).

Because the Chern number $p\neq 0$ obstructs a globally continuous gauge over
$T^2$, we double the Hilbert space to include two copies with Chern numbers
$+p$ and $-p$:
\begin{equation}
  \tilde{H}^0(\boldsymbol{k})
    = \begin{pmatrix} H_p^0(\boldsymbol{k}) & 0 \\ 0 & H_{-p}^0(\boldsymbol{k}) \end{pmatrix}.
\end{equation}
The doubled system has net Chern number zero, admitting a continuous gauge.
Denoting the Berry connection of $\tilde{H}^0$ as $\tilde{A}^0$ and inserting
the gauge transformation into the Chern-Simons form $Q_3 =
\mathrm{Tr}(A\,dA + \tfrac{2}{3}A^3)$, one obtains
\begin{equation}
  Q_3 = \mathrm{Tr}\!\left(2Q\tilde{F}^0 - d(Q\tilde{A}^0)\right)\wedge d\phi,
\end{equation}
where $Q_{ij}(\boldsymbol{k}) =
\langle\tilde{u}_i^0(\boldsymbol{k})|\,q\,|\tilde{u}_j^0(\boldsymbol{k})\rangle$
and $q = \tfrac{1+\tau_z}{2}$ projects onto the $+p$ copy.  The total
derivative integrates to zero on the closed manifold $T^2\times S^1$.
The remaining term projects out the Berry curvature $F^0$ of $H_p^0$, so that
\begin{equation}
  Q_3 = \mathrm{Tr}(F^0)\wedge d\phi.
\end{equation}
Integrating over $T^2\times S^1$ yields
\begin{equation}
  \nu = p\cdot m\;\;\mathrm{mod}\;2,
\end{equation}
where $m$ is the total phase winding accumulated around all vacancies.  For
the Haldane model $p=C=\pm 1$, and $m=N_A-N_B$, recovering Eq.~\eqref{eq:z2}.

\bibliography{general}

\end{document}